# Heterogeneity of Global and Local Connectivity in Spatial Network Structures of World Migration

Valentin Danchev[a,b*] and Mason A. Porter[c, d]

[a] Oxford Department of International Development, University of Oxford, 3 Mansfield Rd, Oxford OX1 3TB, UK
[b] Department of Sociology, University of Chicago, 1126 East 59th Street, Chicago, IL 60637, USA
[c] Oxford Centre for Industrial and Applied Mathematics, Mathematical Institute, University of Oxford, Oxford OX2 6GG, UK
[d] CABDyN Complexity Centre, University of Oxford, Oxford OX1 1HP, UK
[*] Corresponding author: vdanchev@uchicago.edu

## Abstract

We examine world migration as a social-spatial network of countries connected via movements of people. We assess how multilateral migratory relationships at global, regional, and local scales coexist ("glocalization"), divide ("polarization"), or form an interconnected global system ("globalization"). To do this, we decompose the world migration network (WMN) into communities—sets of countries with denser than expected migration connections—and characterize their pattern of local (i.e., intracommunity) and global (i.e., intercommunity) connectivity. We distinguish community signatures—"cave", "biregional", and "bridging"—with distinct migration patterns, spatial network structures, temporal dynamics, and underlying antecedents. Cave communities are tightly-knit, enduring structures that tend to channel local migration between contiguous countries; biregional communities are likely to merge migration between two distinct geographic regions (e.g., North Africa and Europe); and bridging communities have hub-and-spoke structures that tend to emerge dynamically from globe-spanning movements. We find that world migration is neither globally interconnected nor reproduces the geographic boundaries as drawn on a world map but involves a heterogeneous interplay of global and local tendencies in different network regions. We discuss the implications of our results for the understating of variability in today's transnational mobility patterns and migration opportunities across the globe.

Keywords

Spatial networks, Community structure, World migration, Human mobility, Globalization

## 1. Introduction

In the context of today's transnational mobility, world migration can be viewed as a network of cross-border movements of people connecting multiple countries at various—e.g., regional, continental, or global—geographic scales. What is the structure of this network of migration between world countries? What is the interplay between movements at different spatial scales? What tendencies—e.g., global interconnectedness, regional fragmentation, and local heterogeneity— have emerged over the latter part of the twentieth century? Did network structures evolve gradually or transform suddenly? What antecedents could have brought about one or another network structure and what are the implications for the variability of migration opportunities for people around the world? In this paper, we address the above questions through an empirical examination of heterogeneous network structures of world migration using theoretical insights and methods at the intersection of network (Wasserman and Faust, 1994, Newman, 2010) and spatial analysis (Batty, 2005, Barthélemy, 2011, Expert et al., 2011, adams et al., 2012).

We represent world migration as a "social-spatial" network using the Global Bilateral Migration Database (Özden et al., 2011). A social-spatial network is defined as a set of nodes (e.g., societies, organizations, or individuals) located in geographic space that are connected to each other via a set of edges associated with length (and cost) (Barthélemy, 2011, Newman, 2010, Wasserman and Faust, 1994). The world migration network (WMN) is a set of world countries embedded in geographic space. The countries in the network are connected to each other via migration edges of various distances, and an edge represents the number of migrants from a sending country $i$ living in a receiving country $j$ at a particular point of time. The spatial aspect of the WMN comes both from the topographical position of nodes and from the geographic constraints on edges. Although with the advancements in transportation and communication technologies, migration is unlikely to diminish with the increase of distance in a manner predicted by the "inverse-distance rule" (Ravenstein, 1885, Zipf, 1946), the length of migration edges is still associated with a cost (e.g., travel and information costs), so longer-distance migration bears a higher cost (Barthélemy, 2011, Gastner and Newman, 2006). The WMN is directed (i.e., the edges have a direction that represents out- and in-migration) and weighted (i.e., edges have weights that represent, in our study, the volume of migrant stock between countries). The WMN is also temporal network, which we represent as a multilayer network (Mucha et al., 2010, Kivelä et al., 2014). Each layer represents bilateral migration stock between 226 world countries for one of the decades of 1960, 1970, 1980, 1990, and 2000.

Three alternative patterns of world migration and associated mobility opportunities for people are articulated in the literatures on international migration and globalization. First, a growing body of migration literature (International Organization for Migration, 2003, Castles and Miller, 2009, Audebert and Doraï, 2010), supported by recent network studies (Davis et al., 2013, Fagiolo and Mastrorillo, 2013), argued that post-1970 international migration has become more interconnected than it was before in response to increasing globalization. However, reports have pointed to a parallel tendency towards migration regionalization, estimating that about 80 percent of the movements in the developing world are between contiguous countries (e.g., Bangladesh to India) (Ratha and Shaw, 2007, Population Division of the Department of Economic and Social Affairs, 2013). Although one can view regionalization as a "stepping stone" to global integration (Dierks, 2001: 214), obstructed by restrictive migration policies (Hatton and Williamson, 2005), a skeptical view sees it as a sign of polarization between deterritorialized "nomads" that enjoy global long-distance mobility and local "poor" who are locked into geographic regions (Hirst and Thompson, 2000, Bauman, 1998, Wallerstein, 1974, Sassen, 2007). A third possibility, conveyed by the notion of "glocalisation" (Robertson, 1992, Robertson, 1995, Wellman, 2002), reconciles the tension between the idea that local interactions are either declining (i.e., globalization) or subordinated to global processes (i.e., polarization) and suggests instead the possibility of coexistence of globe-spanning interactions and local regional interactions.

Rather than examining a predominant pattern of world migration, we characterize the heterogeneity of world migration by uncovering latent network structures known as "communities" (also called "modules" or "cohesive groups") (Newman and Girvan, 2004, Wasserman and Faust, 1994, Mucha et al., 2010, Porter et al., 2009, Fortunato, 2010). A migration community is a tightly-knit subnetwork of countries with dense migration connection internally (relative to a null model describing random connections) but sparse connections to and from other countries in the network (see Fig. 1).

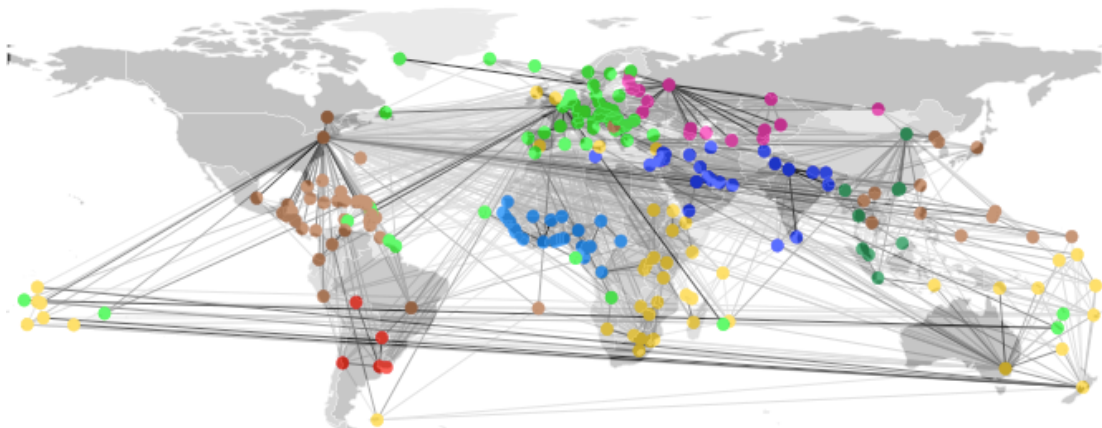

Fig. 1. An example of migration community structure in the WMN. The color of the nodes represents community membership. The position of the nodes indicates the geographic location of countries. The edges represent migratory movements between countries. We represent population size on the map in a gray scale (where darker countries indicate larger populations). We use code from Traud et al. (2009) and Jeub et al. (2015) in MATLAB to visualize the network, and we use the package 'rworldmap' in R (South, 2011) to create the world map in the background.

Detecting communities provides a means of delineating "functional regions" (Ratti et al., 2010) on the basis of empirical connectivity, which may differ from how regional boundaries are drawn on economic and geographic maps (Maoz, 2011: 37). Further, an arrangement of interactions into network communities typically cuts across the hierarchy of spatial scales (Knappett, 2011:

10–11), thus migration communities can encode various combinations of global, regional, and local migration. Third, as Simmel (1950[1908], Martin, 2009, Carrington and Scott, 2011) observed, once cohesive structures crystallize, they can maintain their own existence—and in this way, confront and enable further migration interactions—even if the reasons that brought them to life in the first place have vanished.

To detect migration communities, we employ generalizations of a widely-used method of modularity maximization (Newman and Girvan, 2004) to directed networks (Leicht and Newman, 2008, Arenas et al., 2007), temporal networks (Mucha et al., 2010), and spatial networks (Expert et al., 2011, Sarzynska et al., 2015). Subsequently, we employ statistical techniques to identify significant *migration community signatures*—i.e., communities with similar pattern of global and local migration relationships. We advance prior research (Salt, 1989, Nogle, 1994, DeWaard et al., 2012, Fagiolo and Mastrorillo, 2013) by extracting mesoscale structures in world migration in a way that simultaneously takes into consideration directionality, temporal dynamics, and geographic constraints on migration relationships.

Our contribution extends beyond a descriptive portrayal of typologically different spatial network signatures of glocal migration patterns, as we also show that different migration community signatures are associated with distinct temporal dynamics and relational, social, and spatial antecedents, thereby having distinct implications for migration opportunities across the world. Our analytical framework enables an examination (and reconciliation) of conflicting views of global processes, and our findings may consequently contribute to the understanding of broader transnational interactions.

Traditionally, research on international migration (and other cross-border relationships, such as international relations and trade), has considered each migratory movement between a dyad of countries as independent, attributing variations in migration outcomes to differences in the characteristics of origin-destination pairs (Kim and Skvoretz, 2010, Lupu and Traag, 2013, Fagiolo and Mastrorillo, 2013). The dyadic-independence assumption appears to have been largely applicable to the post-World War II bilateral migration until the 1970s as, in Vertovec's (2010: 3–4) words, 'large numbers [were] moving from particular places to particular places' (e.g., Algeria–France, Turkey–Germany). Since the early 1980s, however, an increasing number of countries have been involved in migration (Audebert and Doraï, 2010: 203, Castles and Miller, 2009: 10), leading to a pattern of 'small numbers moving from many places to many places' (Vertovec, 2010: 3–4). Consequently, one is more likely to observe interactions between multilateral movements of people connecting countries at various geographic— i.e., local, regional, continental, global—scales over time. This increases the probability of extra-dyadic dependencies (Wasserman and Faust, 1994, Newman, 2010) between movements, so a dyadic edge between a pair of countries could depend in part on the patterns of relationships between surrounding—in network and geographic space—countries (Malmberg, 1997).

Prior research in geography (Hägerstrand, 1957, Kritz et al., 1992, Fotheringham, 1991), and more recently in network analysis, of internal (Maier and Vyborny, 2008) and international migration (Nogle, 1994, Davis et al., 2013, Tranos et al., 2012, Fagiolo and Mastrorillo, 2013) has examined extra-dyadic and meso-scale properties of migratory movements. Other studies have examined network properties of global migration in relation to global networks of short-term human mobility (Belyi et al., 2016), international trade (Fagiolo and Mastrorillo, 2014), and international flights and digital communication (Hristova et al., 2016). By integrating network and spatial considerations[1] in the context of globalization theories, we wish to extend past research through an examination of spatial network structures that emerge from multilateral and multiscale movements of people, the way these structures are shaped by relational, social, and spatial antecedents, and the way in which, in turn, they distinctively shape migration opportunities across the globe.

---

[1] For an example of such integration in research on internal migration, see Lemercier and Rosental's study (2010) of migration between villages in 19th century Northern France.

## 2. Structure, Dynamics, and Antecedents of the WMN

Network analysis focuses on relationships between entities (e.g., people, institutions, or countries) in an interconnected system. It provides a vantage point from which one can capture macro-scale and meso-scale network structures that emerge from such relationships (Wellman and Berkowitz, 1988, Wasserman and Faust, 1994, Newman, 2010, Maoz, 2011) as well as sources of opportunities and constraints pertaining to one or another structure (Borgatti et al., 2009). A basic premise of our work is that world migration is an instance of an interconnected system. This system involves multiple cross-border movements of people that connect geographically dispersed locations, which give rise to enduring multilateral migration structures. What structures of world migration have emerged over the second half of the twentieth century?

### 2.1. Globalization, Polarization, and Glocalization

Globalization theories (Beck, 2000, Giddens, 1990, Held et al., 1999) have emphasized the intensification of transnational interconnectedness across the globe, such that distant societies have been integrated in networks of relationships. As Sassen (2007: 137) argued, the increase of foreign direct investments and the export of manufacturing activities to developing areas (e.g., Asia and South America) have contributed to a densification of international capital flows and the formation of new migration pathways. Simultaneously, advancements in transportation and communication technology have shrunk or compressed geographic and cultural distances, a phenomenon known as "time-space compression" (Harvey, 1989). Globalisation is viewed as constituting a 'new historical conjuncture' (Glenn, 2007: 34, McGrew, 1998), in which, as Castells (2010: 440–448 [1996]) put it, the space of places has been replaced by space of flows of capital, goods, information, and people that are spanning across the globe. Under the "globalization of migration" hypothesis, migration scholars have emphasized the progressively increasing number of countries involved in migration (Castles and Miller, 2009: 7–12) and the diversification of origin and destinations since 1970s (Vertovec, 2010), and have observed that major migratory movements are now globe-spanning (e.g., China to the USA) rather than, as in the recent past, exclusively between contiguous countries (e.g., Ireland to England) or bound by past colonial relationships (e.g., Bangladesh to Britain) and bilateral agreements (e.g., between Germany and Turkey) (International Organization for Migration, 2003: 4, Zlotnik, 1998: 465, Agnew, 2009: 170, Castles and Miller, 2009: 7–12, King, 2002: 94).

The world systems theory (Wallerstein, 1974) offers an alternative understanding of globalization. Instead of global integration, the theory has proposed that economic interdependencies have divided the world system into threefold-subordinated strata[2] of core, semi-periphery, and periphery countries. A central tenet of the theory is that peripheral countries are disadvantaged not because of exclusion but because of their assimilation to the world system in a "structurally subordinate position" (Boli and Lechner, 2009: 326). Because of the dependence on restrictive policies of core countries, Hirst and Thompson (2000) argued, current globalization reduced opportunities for international mobility. According to Hirst and Thompson (2000), in the *belle époque* for 1890 to 1914, due to the "border openness" and the "empty lands", flows of goods, investment capital, and labour migration were comparable in magnitude to or even greater than those in the latter half of the twentieth century. In this account, globalisation has neither flattened world stratification nor contributed to an overall increase in economic and mobility possibilities but widened the gap between rich countries and poor countries (Sassen, 1988, Hirst and Thompson, 1999), leading to polarization between constraint-free global mobility and local migrations trapped in bounded regions.

The concept of glocalization conveys the idea of global and local tendencies as simultaneously present and mutually reinforcing (Robertson, 1992, 1995). When defined in spatial terms, glocalisation refers to the coexistence of dense local connections and sparse global—i.e., long-distance—connections (Wellman, 2002). The concept of glocalization offers an alternative to the understanding of globalization as a linear process towards global integration. Also, while the world systems theory (Wallerstein, 1974) emphasizes the differential distribution of movements across regions, glocalization highlights the possibility of a country to simultaneously maintain dense migration connections to neighbouring countries and disperse connections between continents.[3]

---

[2] In Skeldon's (1997) account, each stratum in the world system induces specific migration patterns.

## 2.2. Global and Local Cohesion

To characterize the threefold patterns of interplay—globalization, polarization, glocalization—between global and local migration connectivity in world migration communities, we employ a community-scale conceptualization of Granovetter's strength-of-weak-ties hypothesis (Granovetter, 1973: 1373) that was recently rejuvenated in the theoretical work of Borgatti and Lopez-Kidwell (2011: 42). In the context of international migration, the hypothesis states that if countries $i$ and $j$ share a stronger migration edge (measured in terms of number of migrants), their neighborhoods are more likely to overlap. In other words, they are more likely to have migration connections to the same third countries, leading to strong local (intracommunity) cohesion. In contrast, weak migration edges are likely to perform a bridging role, thereby connecting countries with otherwise disjoined neighborhoods, contributing to strong global (intercommunity) cohesion. This framework allows us to arrange world migration communities on a continuum from communities with strong local (intracommunity) cohesion but weak global (intercommunity) cohesion and communities with weak local cohesion but strong global cohesion (see Fig. 2). We use the distribution of global and local cohesion to establish statistically significant migration community signatures.

---

[3] The coexistence of global and local (regional) migratory movements is discussed in Held et al. (1999) and was recently adopted in different contexts, for example, international trade (Zhu et al., 2014).

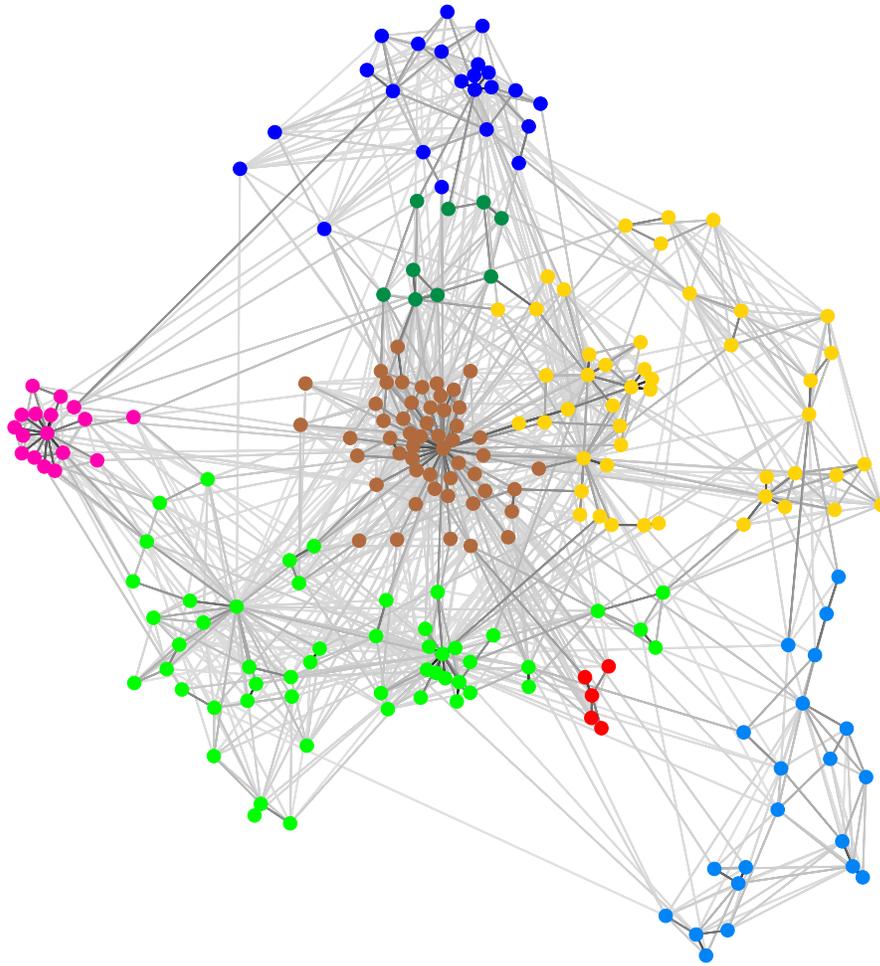

Fig. 2. An example of migration communities that differ in local and global connectivity. The 226 nodes in the WMN are assigned to one of eight different communities. For visual purposes, we symmetrized and thresholded edges (darker edges indicate larger migratory movements). The brown community in the center of the network exemplifies strong global cohesion, whereas the blue community in the lower right corner exemplifies relatively strong local cohesion. We use code from Traud et al. (2009) and Jeub et al. (2015) in MATLAB to create the visualization.

## 2.3. Community Evolution

Compared to more transient forms of global human mobility, such as tourism (Belyi et al., 2016), migratory movements are characterized by an enduring pattern due to their tendency to self-perpetuate via migrant networks (Hägerstrand, 1957, Palloni et al., 2001). This observation is consistent with the world systems theory, which sees nothing unprecedented in the post-1945 migratory movements compared to the early decades of that century. By contrast, globalization theories predict a marked change in global migration since 1970s. We study the evolution of world migration communities by tracking changes in their membership composition.

## 2.4. Relational, Homophily, and Spatial Antecedents

To account for the variability across migration community signatures, we consider a set of endogenous network, homophily, spatial, and economic mechanisms. We first consider local (e.g., dyadic, triadic) mechanisms, such as reciprocity. Reciprocity refers to the tendency of an edge from node $i$ to node $j$ to be accompanied by an edge in the opposite direction (Wasserman and Faust, 1994, Butts, 2008). The tendency has been known in migration studies since Ravenstein (1885:199), who stated that '[e]ach main current of migration produces a compensating counter-current'. A related mechanism—triadic closure—refers to the tendency for an edge to occur between nodes $i$ and $j$ if they are already connected to a common third node $k$ (Davis, 1967, Wasserman and Faust, 1994, Easley and Kleinberg, 2010, Granovetter, 1973). Reasons for triadic closure in world migration can include exogenous forces such as homophily and geographic proximity as well as mechanisms that are generated in the process of migration—e.g., information transmission, migration policies, and migrant networks—and feed back into the WMN. Greater reciprocity and triadic closure typically contribute to strong local cohesion.

    A mechanism that can contribute to global cohesion in networks is hub formation. Hubs are countries that are disproportionately well-connected (Newman, 2010: 245, Slater, 2008) and could integrate a set of spokes. Hub-and-spoke structures are likely to have an uneven distribution of connectivity of well-connected core 'hubs' and sparsely connected peripheral 'spokes' (Borgatti and Everett, 2000, Rombach et al., 2014). Hub-and-spoke migration structures could emerge endogenously as a function of cumulative advantage (de Solla Price, 1965). In international migration, cumulative advantage would suggest that already popular destinations are likely to attract more migrants from diverse destinations. Exogenous forces for hub-and-spoke structures include economic attractiveness and time-space compression.

Reciprocity and triadic closure are more likely to occur if the societies in question are similar along relevant characteristics. This is an example of "homophily" (McPherson et al., 2001, Lazarsfeld and Merton, 1954, Moody, 2009). We consider two homophily mechanisms: former colonial[4] relationships and language proximity (Fawcett, 1989, Portes and Böröcz, 1989, Pedersen et al., 2008, Mayda, 2010, Kim and Cohen, 2010, Breunig et al., 2012). In the context of world migration, homophily tendencies can facilitate local connectivity, and they can also form the basis for long-distance, cross-continental connections (e.g., the Commonwealths).

Once migration exchanges are initiated via homophily mechanisms or other mechanisms (e.g., bilateral agreements), they tend to self-perpetuate over time as a function of social processes that emerge in the course of migration (Massey et al., 1998: 42, Portes and Böröcz, 1989: 612, Massey, 1990). One such social process is "chain migration", in which initial movers are followed by extended family, friends, and acquaintances who obtain access to resources and information about the destination in question through migrant networks (MacDonald and MacDonald, 1964, Gurak and Caces, 1992, Boyd, 1989, Massey et al., 1998). Chain migration can contribute to the formation of hubs in the WMN by channeling migratory movements to particular—and often long-distance—destinations that are typically more costly and risky.

We consider Gross Domestic Product (GDP) per capita as an indicator of economic prosperity. If processes of globalization are relatively evenly distributed, we should expect countries to increasingly involve long-distance movements regardless of their GDP per capita. However, if patterns of migratory movements reflect economic disparities, this would imply polarization.

Migration exchanges can be affected by one or multiple constraints. For example, Martin (2009: 32–36) put forth the hypothesis that strong ties (e.g., friendship) are more likely to require proximity in both geographic space and homophily space. In contrast, weak ties (e.g., acquaintances), according to Martin (ibid. 36), follow an either/or logic, so two actors are likely to know each other if they are affected by either geographic space or homophily space. In the context of the WMN, where strong and weak ties refer to large and small number of migrants, respectively, Martin's hypothesis would imply that a migration edge between a pair of countries is likely to be stronger if the countries are affected by both geographic space and homophily (e.g., similar language) space and weaker if they are affected by only one of the two. The hypothesis also suggests that communities with strong local cohesion are more likely to be induced by both spaces, whereas communities with strong global cohesion are likely to be independent of one of them.

---

[4] Former colonial relationships should be considered only as a proxy of possible social, institutional, and/or cultural similarities. These similarities may have or may have not materialized, and may vanish over time. There is a further complication as former colonial relationships can be also viewed as a network variable (i.e., past relationships between a set of countries).

### 3. Research hypotheses

We formulate below three hypotheses that guide our empirical investigation.

*Globalization hypothesis: An increasing amount of global long-distance migratory movement has contributed to a more interconnected WMN.* The globalization hypothesis implies an unprecedented change in the structure of the WMN, measured in terms of intercommunity density, as well as decoupling from spatial and/or homophily constraints.

*Polarization hypothesis: Global migratory movements have developed in a relative separation from local movements, contributing to a polarized WMN.* The hypothesis implies that communities of local cohesion have not disappeared but are subordinated to communities of global cohesion. Local movements are likely to encounter multiple constraints, compared to relatively constraint-free global movements.

*Glocalization hypothesis: The co-presence of global and local migration in communities is likely to result in a glocal WMN.* The hypothesis implies relatively

even distribution of global and local cohesion across the WMN over time as well as of constraining antecedents.

## 4. Methods, Diagnostics, and Data

In this section, we outline the methods, diagnostics, and data we employ to detect migration communities, characterize their structure, and account for their possible antecedents.

### 4.1. Community Detection

To extend modularity to networks that change over time, Mucha et al. (2010) developed a modularity function for multilayer networks, including ones that represent temporal networks.[5] In the context of temporal networks, the layers are ordered, and it is common to include inter-layer edges only between contiguous layers (Kivelä et al., 2014: 15). To regulate the strength of coupling between time layers, the multilayer modularity function incorporates a temporal resolution parameter $\omega$ (Bassett et al., 2013). By varying the values of $w_{jlr}$, the strength of the connection between node $j$ in layer $l$ (i.e., at time $t_n$) and itself in layer $r$ (i.e., at time $t_{n+1}$) changes. When $w_{jlr} = 0$, each layer is independent as in the static version of modularity. As $w_{jlr}$ is increased, nodes have a stronger incentive to belong to the same community in $t_n$ as in $t_{n-1}$. Communities are then likely to merge across temporal layers, particularly if empirical connectivity changes little over time. The formula for multilayer modularity (Mucha et al., 2010, Bassett et al., 2013) is

$$Q_{\text{multilayer}} = \frac{1}{2\mu} \sum_{ijlr} \{(W_{ijl} - \gamma_l P_{ijl})\delta_{lr} + \delta_{ij}\omega_{jlr}\}\delta(g_{il}, g_{jr}), \quad (1)$$

---

[5] The properties of multilayer modularity were studied further in Bassett et al. (2013) and Bazzi et al. (2016).

where $g_{il}$ is the community of node $i$ in layer $l$ (and $g_{jr}$ is the community of node $j$ in layer $r$), the Kronecker delta $\delta(g_{il}, g_{jr}) = 1$ if vertices $i$ and $j$ are placed in the same community in layer $l$ and layer $r$ and $\delta(g_{il}, g_{jr}) = 0$ otherwise, $\omega_{jlr}$ is the interlayer coupling used to control the strength of the connection between node $j$ in layer $r$ and node $j$ in layer $l$, the quantity $W_{ijl}$ is the element of the weighted adjacency array of layer $l$, the null model $P_{ijl}$ is the expected connectivity in layer $l$, the quantity $\gamma_l$ is the intralayer structural resolution parameter for layer $l$, and $\mu = \frac{1}{2}\sum_{jr} k_{jr}$ is the total edge weight in the network and is a normalization factor that allows the modularity score $Q$ of a partition of a network lies in the range between −1 (all edges are outside communities) to 1 (all edges are within communities). By considering connectivity across temporal layers, multilayer modularity helps capture dynamics that are obscured when temporal networks are represented as a sequence of static snapshots (Mucha et al., 2010).

To factor out statistically unsurprising connectivity, one can implement different null models, depending on what constraints are hypothesized to have an effect on community formation (Newman, 2012, Expert et al., 2011). As edge directionality is an essential feature of the WMN, we employ a modularity null model for directed networks (Leicht and Newman, 2008, Arenas et al., 2007, Malliaros and Vazirgiannis, 2013). Using this null model, modularity estimates whether a partition has more edge weights within communities than expected in an associated empirical network with the same out- and in-strength sequence but with edge weights distributed at random (Newman, 2006). The Leicht–Newman (LN) null model (2008, Arenas et al., 2007) for directed networks is

$$P_{ij}^{LN} = \frac{s_i^{\text{out}} s_j^{\text{in}}}{w}, \qquad (2)$$

where $s_i^{\text{out}}$ and $s_j^{\text{in}}$ are out- and in-strength of node $i$ and node $j$, and $w$ denotes the total weight in the network.

Similar to other spatial networks (Barthélemy, 2011), the nodes and the edges in the WMN have a location and cost, respectively. Therefore, any two countries with similar out-migration strength sequences can have rather different probabilities to connect to a third country as a function of their location in geographic space. To account for the constraining role of geographic attributes, Expert et al. (2011) developed a spatial null model. When extended to directed networks, Expert et al.'s (2011) model is

$$P_{ij}^{\text{Spa}} = N_i^{\text{out}} N_j^{\text{in}} f(d_{ij}), \quad (3)$$

where $P_{ij}^{\text{Spa}}$ is the expected migration stock between country $i$ and $j$, the quantity $N_i^{\text{out}}$ and $N_i^{\text{in}}$ measures the potential of origin $i$ and the attractiveness of destination $j$ (we use the total out- and in-migration for each 226 countries as an indicator for potential and attractiveness), and the 'deterrence function' $f(d_{ij})$ measures the effect of distance. We compute the great-circle geographic distance between the capital cities of the 226 world countries (Furrer et al., 2013). As in gravity models (Anderson, 2011, Haynes and Fotheringham, 1984), the intuition behind Expert et al.'s spatial null model is that $N_i^{\text{out}}$ and $N_i^{\text{in}}$ are sources of opportunities (e.g., possible interactions between a pair of countries), and the distance $d_{ij}$ is a source of constraints. Expert et al. (2011) proposed the following deterrence function:

$$f(d) = \frac{\sum_{\{i,j|d_{ij}=d\}} A_{ij}}{\sum_{\{i,j|d_{ij}=d\}} N_i^{\text{out}} N_j^{\text{in}}}. \quad (4)$$

In the context of the WMN, the deterrence function $f(d)$ is the weighted average of the probability $\frac{A_{ij}}{N_i^{\text{out}} N_j^{\text{in}}}$ for a migration edge weight to exist from country $i$ to country $j$ at a certain distance range. The deterrence function uses bins to calculate the expected migration for a certain distance range. After examination of alternative values, we set the bin size to 500 km.

A larger positive value for spatial modularity $Q^{\text{Spa}}$ indicates a higher density of edge weights inside communities than one would expect for the given null model. Because this spatial null model is designed to allocate a larger contribution to edges between distant nodes than to edges between nearby nodes, the model tries to "factor out" spatial dependence in its detection of communities in the WMN in the hope of shedding light on the role that non-spatial mechanisms (e.g., homophily) can play in their formation.

Modularity maximization is NP-hard (Brandes et al., 2007), and it also has some well-studied limitations, such a resolution limit and extreme near-degeneracy among local maxima with high modularities. The former limitation refers to the tendency of the modularity function to overlook communities that are smaller than some characteristic size (Fortunato and Barthelemy, 2007), although one can ameliorate the issue by incorporating of a resolution parameter $\gamma$ in the modularity function (Porter et al., 2009, Reichardt and Bornholdt, 2006). The latter issue refers to the numerous near degeneracies in the rugged landscape of the modularity function, and partitions with similar high-modularity scores can arise from rather dissimilar structures (Good et al., 2010). To take into account near-degeneracies in the modularity landscape, we identify consensus partitions (Lancichinetti and Fortunato, 2012, Bassett et al., 2013, Bazzi et al., 2016, Sarzynska et al., 2015) across multiple optimizations (see Appendix A). The consensus partitions are robust to variation across optimizations, thereby ameliorating the issue of near-degeneracy. We optimize modularity using the generalized Louvain heuristic (Blondel et al., 2008, Jutla et al., 2011–2012).

## 4.2. Diagnostics for Characterizing Migration Communities

*E-I Index for Weighted Networks*

The E-I index is a widely used measure of group embeddedness (Krackhardt and Stern, 1988, Hanneman and Riddle, 2011: 348, Borgatti et al., 2002) that we use to give a simple measure for the extent to which a migration community exhibits local (intracommunity edge strengths) and global (intercommunity edge strengths) cohesion. We generalized the E-I index to weighted networks

$$\text{E–I index} = \frac{EW - IW}{EW + IW}. \tag{5}$$

The index compares the amount of internal weights $IW$ to the amount of external weights $EW$. The E-I index takes values between $-1$ to $+1$. As the value of the E-I index approaches $-1$, most edge weights are internal to migration communities. As the index approaches $+1$, most edge weights are external to the communities. We apply the diagnostic to both the whole WMN and to each community individually.

*Neighborhood Overlap*

As Onnela et al. (2007) and Easley and Kleinberg (2010: 52) observed, the original strength-of-weak-ties hypothesis (Granovetter, 1973) imposes "sharp dichotomies"— edges can be either strong or weak and can either be local bridges or not bridges. They argued in favor of a continuous definition that captures the gradation in real-world data. They thus defined the neighborhood overlap $O$ of an edge between nodes $i$ and $j$ as the ratio of the number of neighbors that nodes $i$ and $j$ have in common to the number of neighbors of either $i$ or $j$ (Onnela et al., 2007: 7334, Easley and Kleinberg, 2010: 52). The neighborhood overlap $O_{ij}$ ranges from 0 (edges that could serve as bridges between distinct communities) to 1 (edges that connect nodes with overlapping set of neighbors).

*Community Change*

To examine community evolution over time, we compute a temporal autocorrelation function $C(t)$, which quantifies the overlap of community structure at time $t_n$ with itself at time $t_{n+1}$ (Palla et al., 2007). The community autocorrelation is

$$C(t) \equiv \frac{|A(t_n) \cap A(t_n + t)|}{|A(t_n) \cup A(t_n + t)|}, \tag{6}$$

where the numerator $|A(t_n) \cap A(t_n + t)|$ is the number of nodes that belong to both community $A(t_n)$ and community $A(t_n + t)$, and the denominator $|A(t_n) \cup A(t_n + t)|$ gives the number of nodes that belong to at least one of the two communities. The output ranges from 0 to 1, where 0 indicates complete change in the community membership from $t_n$ to $t_{n+1}$, and a score of 1 indicates that a community is identical at times $t_n$ and $t_{n+1}$.

4.3. Antecedents Shaping the Migration Community Signatures

We operationalize below the set variables we use as input in Principal component analysis (PCA) (Jolliffe, 2002) and ANOVA to examine the differential impact of relational, homophily, and spatial mechanisms on the migration community signatures. To account for network dependencies and examine the relative importance of migration antecedents, we employ the multiple regression quadratic assignment procedure (MR-QAP) (Krackardt, 1987, Dekker et al., 2007). We provide details about MR-QAP in Appendix B.

*Reciprocity*

We define reciprocity $R$ in the directed WMN as the fraction of reciprocated edges $M/(M + \frac{A}{2})$, where $M$ denotes mutual edges and A denotes asymmetric edges (Butts, 2008: 27). The reciprocity score for a network ranges between 0 (none of the edges is reciprocated) to 1 (all edges are reciprocated). We consider as reciprocated pairs of edges in the weighted WMN that have migration ratio between them above the threshold of .5.

*Weighted Clustering Coefficient*

We employ weighted clustering coefficient to operationalize the concept of triadic closure. Over the last decade, several papers have generalised the topological formulations of clustering coefficient to weighted networks (Saramäki et al., 2007, Barrat et al., 2004). Onnela et al. (2005) extended the original local clustering coefficient $C_i = \frac{2t_i}{k_i(k_i-1)}$ (Watts and Strogatz, 1998: 201, Newman, 2010) to weighted networks by replacing the number of triangles $t_i$ attached to a node with the sum of triangle weight intensities, yielding the expression

$$\widetilde{C}_i = \frac{2}{k_i(k_i-1)} \sum_{j,k} (\widetilde{w}_{ij}\widetilde{w}_{jk}\widetilde{w}_{ki})^{1/3}, \tag{7}$$

where $k_i$ is the degree of node $i$ and the weight intensities $\widetilde{w}_{ij}$ are normalised values, obtained by dividing weights $w_{ij}$ by the maximum weight $\max(w_{ij})$ in a network. The contribution of each triangle is a function of all of its constituting edge weights, such that triangles with heterogeneous weights will have smaller contributions than triangles with balanced weights (Saramäki et al., 2007: 2). In our calculation, we employ a generalisation of Onnela et al.'s (2005) weighted clustering coefficient to directed networks (Fagiolo, 2007, Rubinov and Sporns, 2010). We compute global weighted clustering coefficient $C_w = \frac{1}{n}\sum_{i=1}^{n}\widetilde{C}_i$ as the mean of the local clustering coefficients of all countries assigned to the respective community.

*Strength Inequality via Gini Coefficient*

To examine the inequality of the distribution of migration strengths in a community, we calculate the Gini coefficient (Kunegis and Preusse, 2012). To compute the Gini coefficient $G_s$ of community $c$, we sort the sequence of migration strengths $s_i$ of all countries in a community. The Gini coefficient is

$$G_s = \frac{2\sum_{i=1}^{n} is_i}{n\sum_{i=1}^{n} s_i} - \frac{n+1}{n}, \ s_i \leq s_i + 1, \tag{8}$$

where $s_i$ is the sorted strength of $i^{th}$ country in a community (Kunegis, 2013). The Gini coefficient takes values between 0 in the case of total equality between migration strengths and 1 in the case of total inequality (i.e., a perfect star community dominated by a single node).

*Community Homophily*

To measure community homophily $H$, we multiply each cell in the binary community matrix of social attributes $S_c$ by the community weighted matrix $W_c$ of international migration for the respective community:

$$H_{ij} = \frac{\sum_{ij} S_{c_{ij}} W_{c_{ij}}}{\sum_{ij} W_{c_{ij}}}, \tag{9}$$

where $S_{c_{ij}} = 1$ when country $i$ and country $j$ share similar language, $S_{c_{ij}} = 0$ otherwise. The measure ranges between 0 (lack of homophily) and 1 (perfect homophily). We apply the measure to language proximity $LP$ and colonial relationships in the past $CRP$. The data come from the CEPII Geodesic Distance Database (Mayer and Zignago, 2006). For language proximity, we created a composite binary variable using two indicators in the database: official language and ethnic language (spoken by at least 9%). A cell in the language proximity matrix is 1 if either country A and B have similar official language or if at least 9% of the population in country A and B speak the same language, and 0 otherwise. For former colonial ties, we create another composite binary matrix, in which a dyad of countries are associated if either they have ever had a colonial link or have had a colonial relationship since 1945.

*Chain Migration*

Given the set of countries included in community $i$ at time $t_n$, we are interested in what proportion of movements follow migration pathways that existed between the same set of countries in $t_{n-1}$. The diagnostic ranges from 0 (none of the migratory movements in community $i$ at $t_n$ follows pathways that existed at $t_{n-1}$) to 1 (all migratory movements in community $i$ at $t_n$ followed pathways that already existed at $t_{n-1}$). Only pathways that involve migration frequencies above the mean for the respective community are considered. In this way, we discard small migration exchanges that do not fall under the definition of chain migration. We do not include chain migration (CM) in the PCA because we have less time points (1960 is only a reference year against which we compute chain migration for 1970).

*Expected Migration Distance*

We simultaneously account for geographic distance and migration stocks by computing the expected distance of a randomly selected migrant for each community. In particular, we weight distance between country $i$ and $j$ by the total number of migrants traveling between country $i$ and $j$ (out- and in-migration), and divide the product by the total in- and out-strength of country $i$:

$$D_i = \frac{\sum_{i=1} w_{ij} d_{ij}}{s_i}, \qquad (10)$$

where $w_{ij}$ denotes the total migration between countries $i$ and $j$, $d_{ij}$ denotes the geographic distance between the two counties, and $s_i$ denotes the strength of country $i$. The community expected migration distance is measured as the average of the expected distance of all countries in community $c_i$, which reads $D_{c_i} = \frac{1}{n}\sum_{i=1}^{n} D_i$. The resulting expected score can also be interpreted as a weighted average distance.

*Economic Disparities*

To account for economic disparities in the WMN, we consider community gross domestic product (GDP) per capita. The source for the data about GDP per capita is from the World Development Indicators (World Bank, 2010).

## 4.4. The Global Migration Database

We construct the WMN from migration stocks for each decade from 1960 to 2000, as recorded in the Global Bilateral Migration Database (Özden et al., 2011). Migrants are defined primarily on the basis of birth country, but other criteria—e.g., country of citizenship—have were also considered (Özden et al., 2011). The database contains comprehensive information about migration stocks (i.e., number of people that were born in country $i$ and lived in country $j$) from national censuses and population registers for 226 countries, resulting in five 226 × 226 matrices. National census surveys are typically carried out at the end of a decade, gathering information about the number of foreign-born people (or foreign citizens) that resided in a given country for at least one year during the preceding decade (UNDESA, 2013). The database reports aggregate migration stock for each decade between 1960 and 2000. Aggregate migration stocks can overlook differences among types of migration (e.g., labour or education) or dynamic forms of migration [e.g., "stepwise" migration (Paul, 2011)], we concur with Bilsborrow and Zlotnik (1994: 66) that in comparison to flow data, migration stocks represent "the long-term effects of migration and [are] thus a more stable component" of international movements. Because of these features, the stock data are instrumental in examining a spatial-network community structure in the WMN.

## 5. Mapping the Landscape of the WMN

The first step of our analysis is to detect migration communities using two different null models, for directed and spatial networks. The models yield different results, which we describe in 5.1 and 5.2, respectively.

## 5.1. Communities Detected via Multilayer LN Modularity

In Fig. 3, we show world maps[6] of consensus community assignments for each decade from 1960 to 2000 obtained using the null model for directed networks.

---

[6] To create the "choropleth" maps, we employ the package 'rworldmap' in R for mapping global data (South, 2011).

We observe eight[7] migration communities in 1960. We label communities with the code of the country that has the largest intracommunity migration strength.

As one can see, geographic distance and regional boundaries play important roles in demarcating the structure of more than half of the migration communities. The geographic signature is imprinted in the communities centered on India (IND), the former Soviet Union (RUS), and China (CHN), as well as in those confined to Sub-Saharan Africa (COD) and West Africa (CIV).

Two migration communities, grouped largely on the basis of ex-colonial relationships, are associated with the principle of homophily: France and countries in North Africa (FRA); and countries in South Europe, South America, and Angola in Africa (ARG). Although homophily is correlated predominantly with geographic proximity in the former community (FRA), the cross-continental grouping between the latter set of countries (ARG) is relatively independent from distance.

We identify cross-continental communities that overcome geographic constraints. This tendency occurs in the largest community in 1960 (USA), which includes North America, Australia, New Zealand, and the bulk of Western, Central, and Northern Europe. This community assignment is fairly unexpected because, in a period that precedes transportation advancements, it groups long-distance migration between non-contiguous countries that are geographically dispersed. In addition to the USA community, the groupings of South America and Mediterranean countries in Europe (ARG) and of North Africa and France in 1960 (FRA) suggest that cross-continental migration may have a statistically significant impact on migration communities. Thus, migration groupings need not be confined to the continental boundaries of the world (e.g., Salt, 1989).

---

[7] The consensus partitions tend to decompose a network into a smaller number of communities (and possibly more consistent) compared to the original partitions.

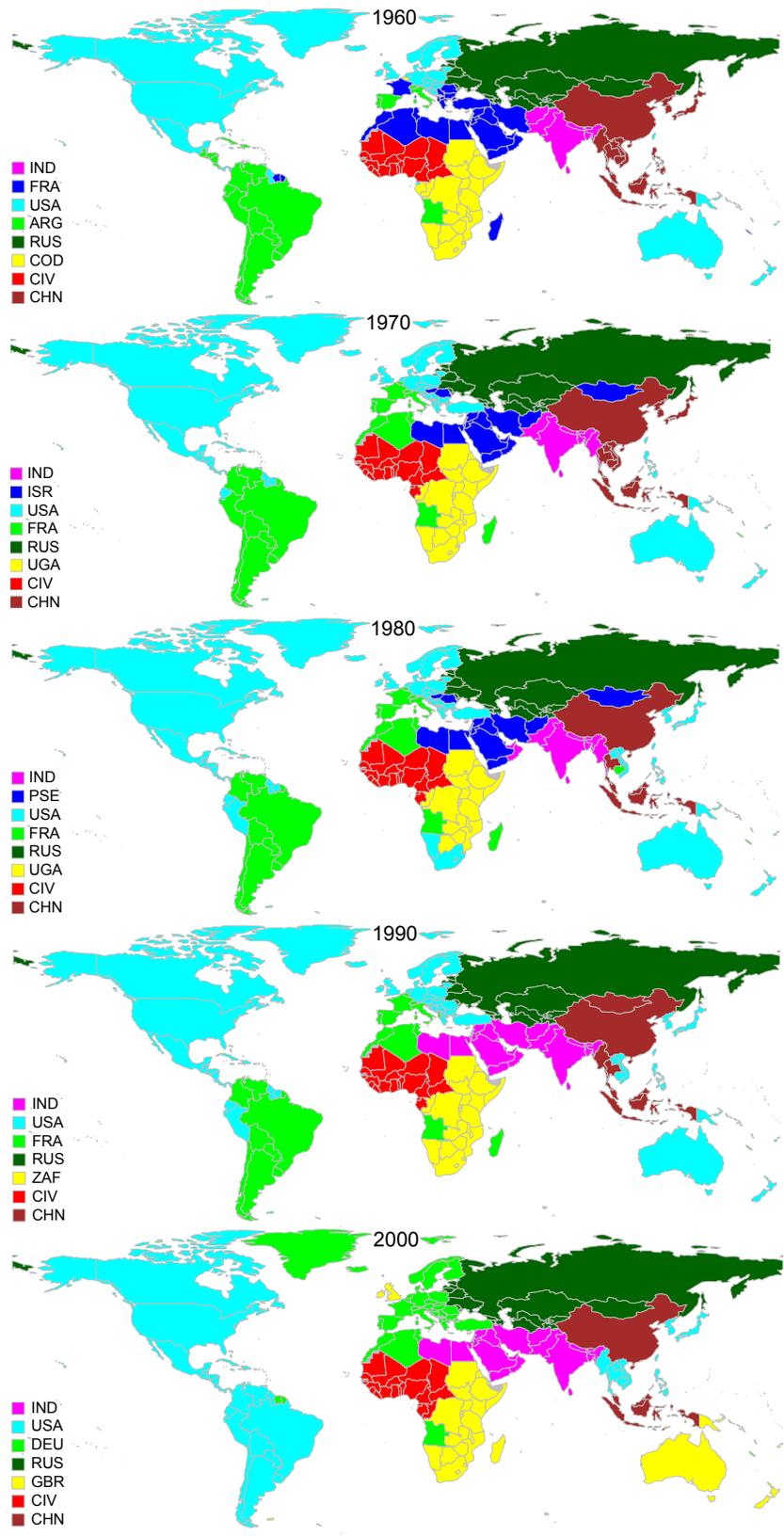

Fig. 3. Migration communities detected via multilayer LN modularity at a resolution of $\gamma = 1$. Color coding indicates community assignments.

Although the aggregate structure of communities has remained virtually intact in the following three decades (1970, 1980, and 1990), one change is worth noting. Since 1970, the global community (USA) involving Western, North, and Central Europe was extended to Eastern Europe and reached Turkey on the South, reflecting the increased migration exchanges between Germany and Turkey following the bilateral recruitment agreement between the two states signed in 1961 (King, 1993).

The community structure of the WMN changed more noticeable in 2000. First, aside from the United Kingdom, all European countries (which were previously separated into two communities) are now assigned to one integrated community (DEU). This result is consistent with Salt's (2001: 3) observation that a characteristic feature of European migration in the middle and late 1990s is '[t]he increasing incorporation of Central and Eastern Europe into the European migration system as a whole'. However, European migration is not separated from other continental 'wholes' but includes North African countries. Therefore, the delineation of the migration map according to geo-political divisions (e.g., the European Union) (e.g., Massey et al., 1998: 110) may not reflect empirical migration connectivity. Second, the United Kingdom, Australia, and New Zealand are no longer part of the largest migration community but instead formed a Commonwealth community that also includes countries in Southeast Africa (GBR). This community exemplifies the role of homophily in connecting geographically dispersed countries.

## 5.2. Communities Detected via Multilayer Spatial Modularity

Although the communities identified via LN modularity reflect some distant homophilous relationships, particularly in year 2000, the communities generated via spatial modularity captures more refined non-spatial structures along time. For example, while Europe appears increasingly integrated over time when using LN modularity, maximizing modularity using the spatial null model suggests the opposite tendency (see Fig. 4). In this case, European migration breaks into a set of small communities, particularly noticeable in the year 2000. This fragmentation pattern exemplifies a key feature of spatial modularity. Because geographic space 'glues' nearby nodes together, the extraction of the effect of geographic distance leads naturally to spatially fragmented communities. The reason for this fragmentation is that migratory movements between nearby countries are likely to have lower contribution to modularity under the spatial null model. By implication, non-contiguous countries have higher probability to be assigned to the same community (e.g., France and Romania are part of the same spatially discontinuous community in 1960, 1970, and 1980).

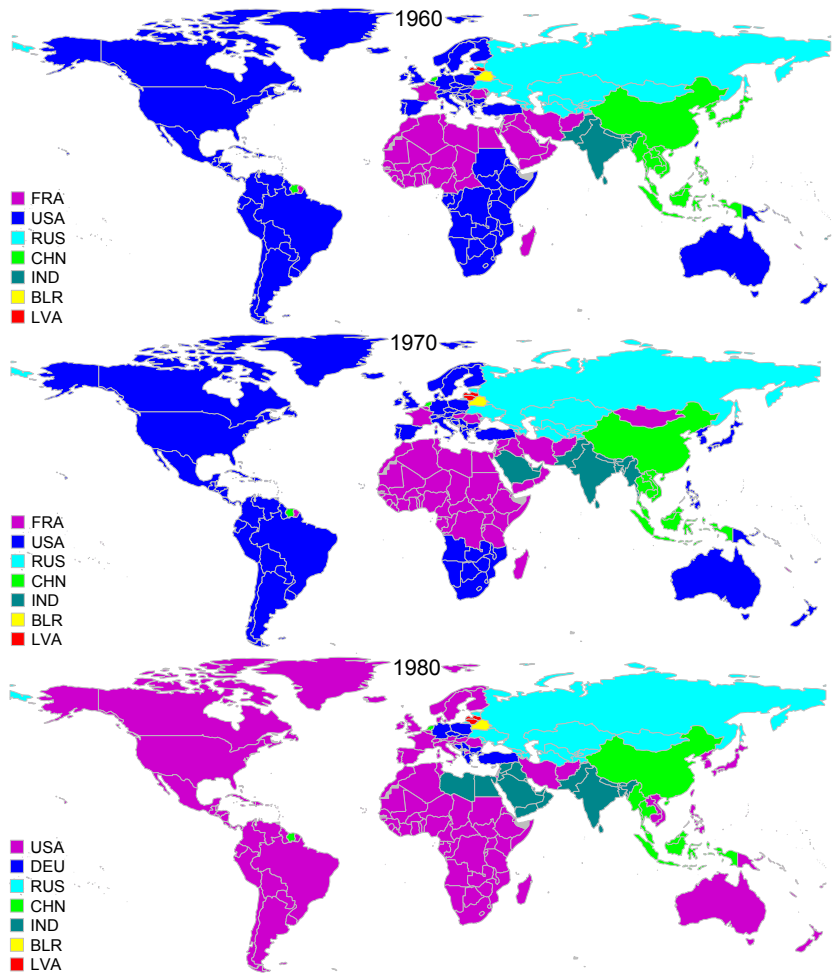

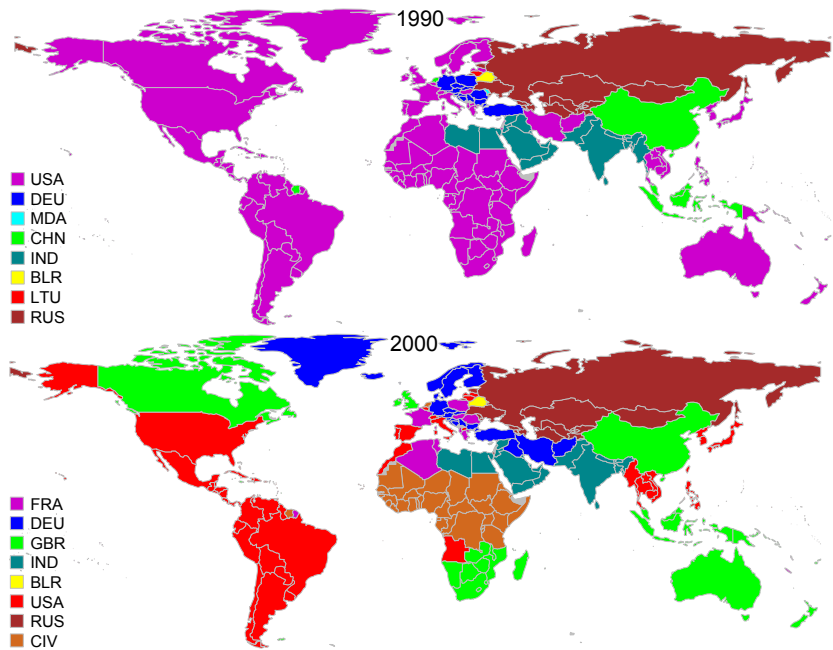

Fig. 4. Migration communities detected by maximizing spatial modularity at a resolution of $\gamma = 1$. Color coding indicates community assignments.

## 5.3. Geographically Contiguous versus Non-contiguous Communities

Much of the previous research on meso-scale groupings in international migration was informed by the migration systems approach (Fawcett, 1989, Kritz et al., 1992). A migration system is defined as 'a group of countries that exchange relatively large numbers of migrants with each other' (Kritz and Zlotnik, 1992: 2). Despite some disagreements about where to draw the boundaries in post-war European migration (Bonifazi, 2008: 123-125), there has been a general consensus on how to draw them. Migration systems were typically viewed as "geographically discrete" [8] (i.e., systems reflect well-delineated geographical areas), consisting of contiguous countries, and confined to the continental boundaries (DeWaard et al., 2012, Salt, 2001, Zlotnik, 1998, Massey et al., 1998: 110). Our findings suggest that (European) movements are not exclusively grouped into geographically distinct regions; rather, such groupings are one empirical possibility among many (including non-contiguous communities). They also suggest that European countries are connected to areas (e.g., North and South Africa) outside of the continent in intercontinental communities that appear to be more stable over time than many intracontinental migration groupings.

---

[8] Most network partitioning methods assign each node to a single community (Porter et al., 2009), and communities are hence 'discrete' in the sense that they do not share nodes with any other community. However, communities need not be discrete geographically (i.e., formed of geographically contiguous countries), and using methods that allow overlapping communities would be an interesting way to further explore such features.

### 5.4. Processes of Fragmentation and Integration in the WMN

The WMN has become marginally more integrated between 1990 and 2000, as reflected in the increasing E-I index computed for the whole network (see Fig. 5). Specifically, the proportion of intercommunity edge weights has increased since 1990, resulting in a more interconnected network. Differences in the network-scale E-I indices over time are more pronounced in the communities detected by maximizing spatial modularity—from $-0.66$ in 1960 to $-0.36$ in 2000—than LN modularity ($-0.50$ in 2000; $-0.59$ in 1960). Our results are consistent with recent findings reported by Davis et al. (2013) and Fagiolo and Mastrorillo (2013). Although the globalization hypothesis receives support at the network scale, the simultaneous presence of global and contiguous communities points to meso-scale heterogeneity that we explore in the following sections.

To examine whether community detection can better specify boundaries between world regions than geographic classifications, we compare our community structures to the macroscale geographical areas (continents) and subregions (e.g., Northern Europe, Southern Europe) that are described in the United Nations (UN) Statistical Division.[9] The E-I indices in Fig. 5 indicate that the communities that we obtain by maximizing modularity with either the LN null model or spatial null model outperform available geographic-based world divisions. By "outperform", we simply mean that, the communities that we identify contain more migration weights within groupings rather than between groupings in comparison to the deterministic geographic-boundary specifications, and in that sense they provide a better way to set boundaries in the WMN.

---

[9] Retrieved on 15 August 2014 from http://unstats.un.org/unsd/methods/m49/m49regin.htm.

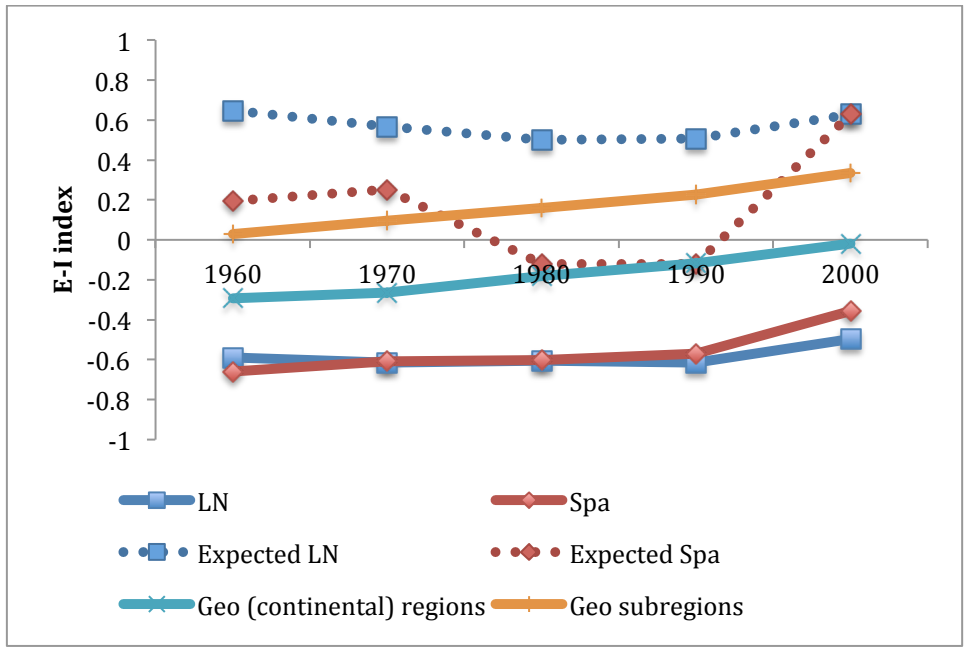

Fig. 5. E-I indices of the weighted WMN. The E-I index characterizes the relationship between intracommunity and intercommunity connectivity in the WMN.

Because the intracommunity and intercommunity connectivity (which are mesoscale properties) can be conditioned on global properties, such as network density, we need a null model to give context to our computation of E-I indices. We perform a permutation test (with 1,000 permutations) and compute the number of times that the observed E-I index is significantly smaller than the expected E-I index measured in an ensemble of null-model WMNs in which rows (out-migration) and columns (in-migration) are simultaneously reshuffled. We find that the observed E-I index is significantly different ($p$-value $< .01$) from the expected E-I index across models (see Fig. 5). The distribution of intracommunity and intercommunity migratory movements is therefore less an artifact of global connectivity but reflect genuine meso-scale patterns of relationships in the WMN.

## 6. Global and Local Cohesion in Migration Communities

To begin to examine the modes of interplay between global and local cohesion—i.e., globalization, polarization, and glocalization—in migration communities, we first generate weighted community adjacency matrices $W_c$ for each decade (see

Fig. 6). We define $W_c$ as follows. For each time point, we sum over all migration edge weights between the 226 countries in the WMN depending on whether an edge remains within community $A$ or lies between a pair of communities $B$ and $C$. In the resulting community adjacency matrices $W_c$, nodes represent migration communities. The edges that remain within communities appear on the main diagonal, and edges between communities appear off of the diagonal. Because the propensity of internal and external connectivity is constrained by the number of communities, their relative size, and edge density (Hanneman and Riddle, 2011), we normalize the community adjacency matrices versus the mean intracommunity and intercommunity edge strengths. In this way, we control for heterogeneity in community size and edge density.

As one can see from Fig. 6, stronger migration edges are more likely to remain within communities than between communities. Moreover, migration communities are differentiated on the basis of their internal edge strength. The communities centered on India, Russia, and China are characterized by relatively high intracommunity edge strength, irrespective of the null models that we employ when maximizing modularity. A different pattern of low intracommunity migration strength is encoded in communities centered on the USA, GBR, and France.

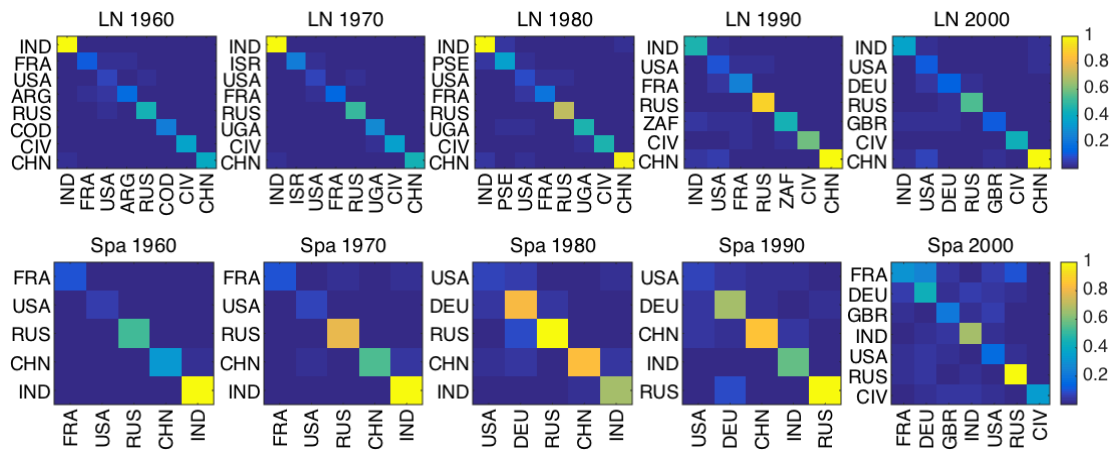

Fig. 6. Normalized community adjacency matrices $W_c$ of intracommunity and intercommunity migration edge strengths. Each element in the matrices represents whether migration edge strength remains within community $A$ or between communities $B$ and $C$. The magnitude of edge strength ranges from weak (in blue) to strong (in yellow). We exclude communities of size $N_c \leq 2$ countries.

We now consider edge neighborhood overlap (see Fig. 7). As expected, migration edges with a large neighborhood overlap are more likely than those with a small overlap to remain within communities than between communities. Instances of large intercommunity neighborhood overlap are rare and typically involve geographically close communities (e.g., India and China, and Uganda and Ivory Coast), indicating a negative relationship between distance and neighborhood overlap. Edge neighborhood overlap is differentially distributed across communities. For example, the communities centered on Russia, Ivory Coast, and India tend to exhibit greater neighborhood overlap than communities associated with the USA across all time layers.

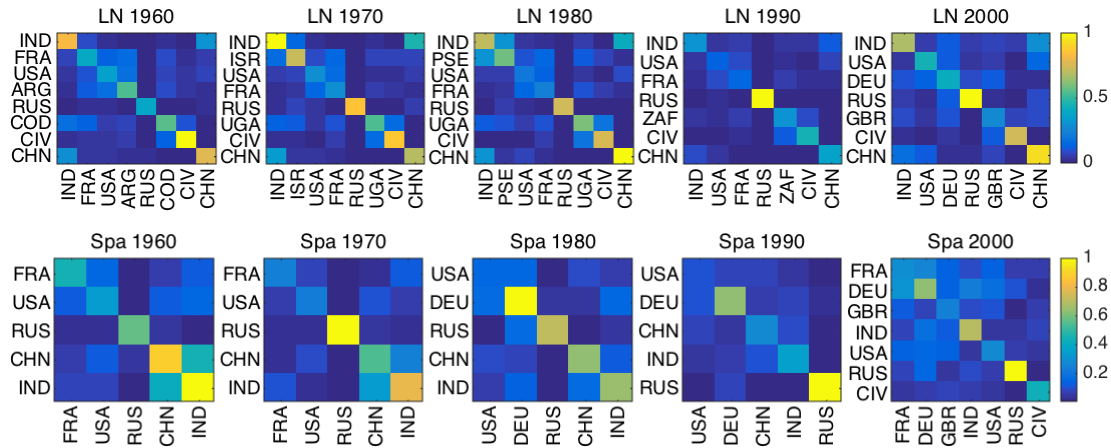

Fig. 7. Normalized community adjacency matrices of intracommunity and intercommunity edge neighborhood overlap.

To begin to identify migration community signatures, we first quantify the relationship between global (intercommunity weak ties) and local cohesion (intracommunity strong ties) by computing E-I indices—i.e., the proportion of external to internal edges—at the community scale for both edge strength ($EI_{es}$) and edge-neighborhood overlap ($EI_{no}$). We then partition the two indices using agglomerative hierarchical clustering[10] in order to identify sets of communities

---

[10] We employ Euclidian distance to determine pairwise (dis)similarities and then average linkage clustering (Newman, 2010: 388, Porter et al., 2009: 1084) to sequentially group communities into a dendrogram (i.e., trees that illustrate a community hierarchy). Using ANOVA tests to evaluate alternative numbers of factions, we find that a three-group partitioning maximizes intergroup variability and minimizes intragroup variability in both $EI_{es}$ ($F_{2,62} \approx 145.37, p < .001$) and $EI_{no}$

with characteristic patterns of distribution of $EI_{es}$ and $EI_{no}$. In Fig. 8, we show two dendrograms for the resulting three-group partitioning of migration communities that we detect by maximizing modularity with the LN and spatial null models. Despite some differences, the typology in the two dendrograms tends to agree: communities centered on the USA, France, Germany, and the United Kingdom are assigned to the left cluster; communities centered on Russia are assigned to the right cluster; and communities associated with India and China are placed in the center cluster.

---

($F_{2,62} \approx 158.58, p < .001$), where F is the ratio of inter-group variance to intra-group variance (see Field, 2009: 359) and the subscripts give the degrees of freedom. We use algorithms implemented in the Statistical Toolbox in MATLAB 2014b.

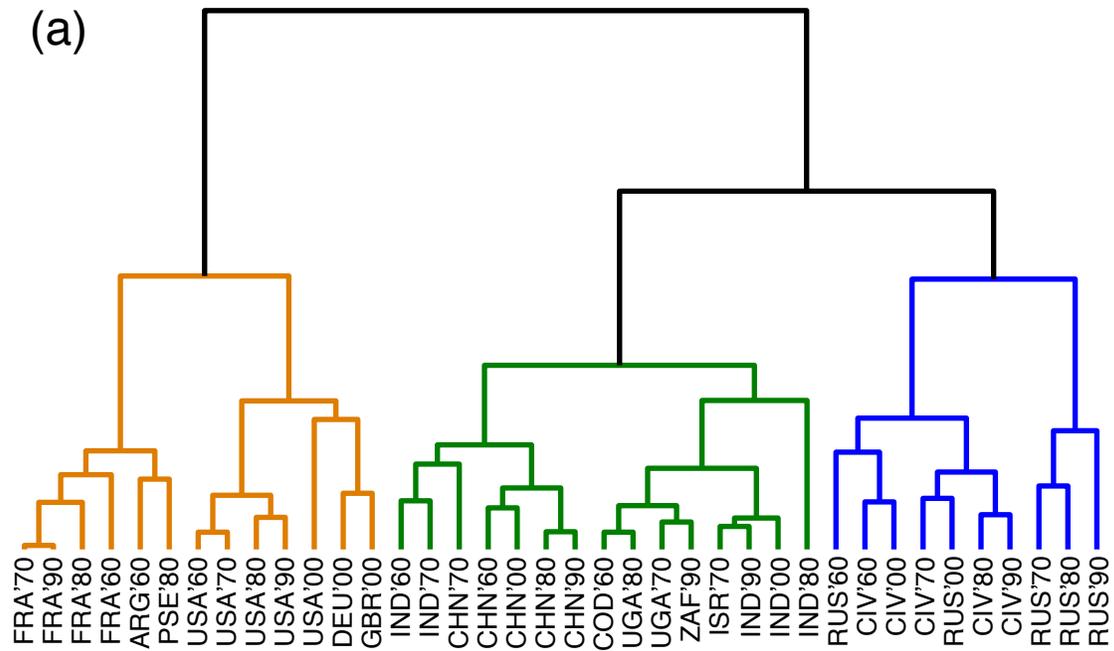

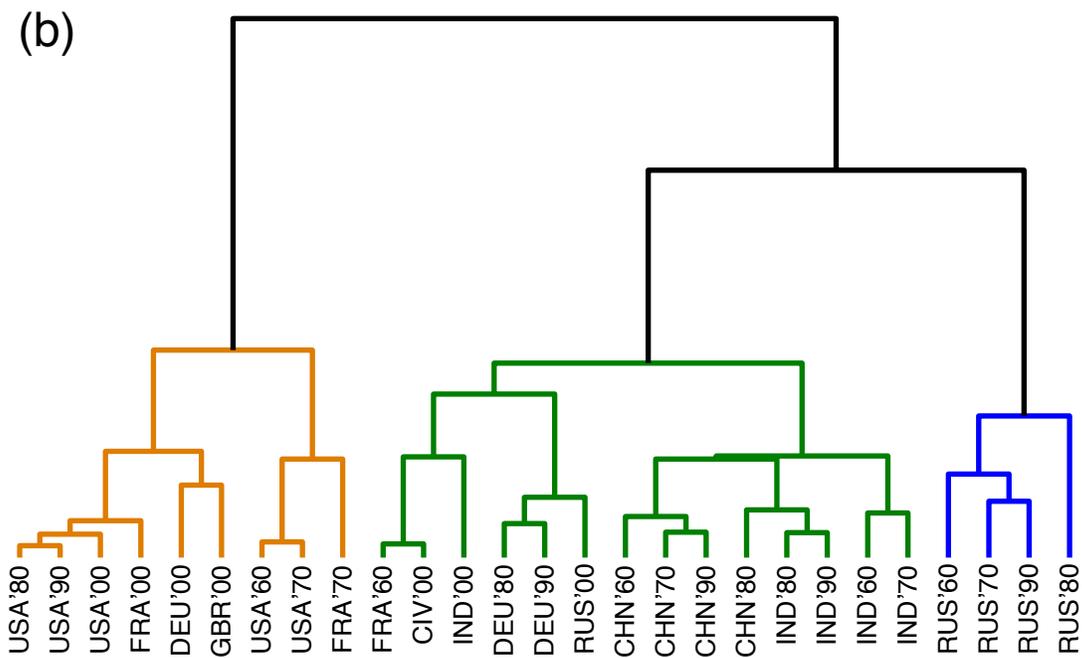

Fig. 8. Dendrogram of migration communities on the basis of E-I edge strength and E-I edge-neighborhood overlap for communities that we obtain by maximizing modularity using (a) LN modularity (38 communities) and (b) spatial modularity (27 communities). The color of the branches represents the three detected factions. We name the communities at the bottom of the dendrogram.

The strength-of-weak-ties hypothesis (Granovetter, 1973) asserts that edge strength should have an impact on neighborhood overlap: a stronger migration connection between a pair of countries increases the likelihood that the two countries will connect to similar third countries, forming a tightly-knit structure. A linear-regression model with an interaction term (between $EI_{es}$ and community signature) indicates (Adj. $R^2 \approx 0.884$) that the variation in $EI_{no}$ is reduced by about 88% when we take into account $EI_{es}$, community cluster, and their interaction (see Fig. 9). The interaction model has more predictive power than a regression model that overlooks community cluster as a covariate ($R^2 \approx 0.566$), suggesting that migration communities are typologically different with regard to their intracommunity and intercommunity cohesion, and these differences mediate the relationship between edge strength and edge-neighborhood overlap.

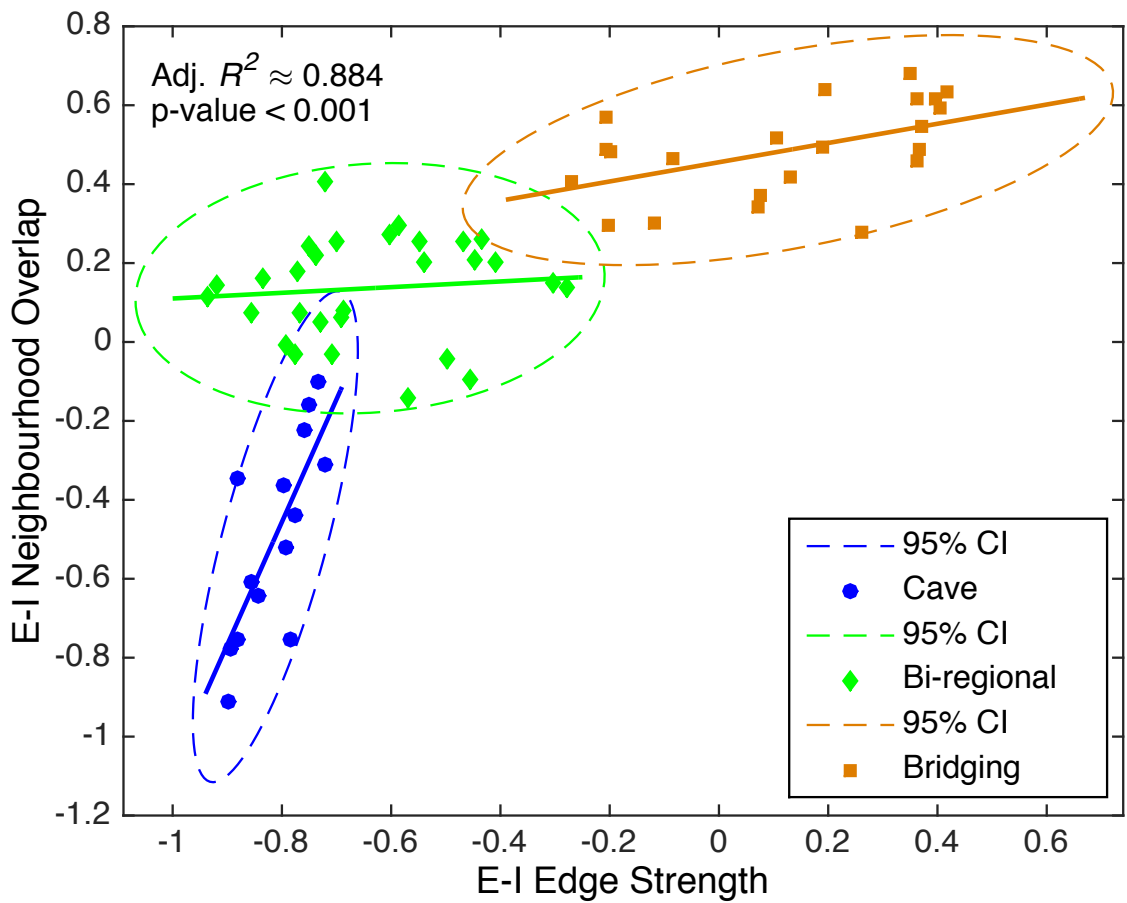

Fig. 9. Relationship between E-I edge strength and E-I edge neighborhood overlap mediated by community signature. We include fitted regression lines for the corresponding community types. The ellipses in dash lines indicate the 95% confidence interval error for the corresponding community signature.

## 6.1. Migration Community Signatures

Using this novel approach to characterizing the structure of the WMN based on intracommunity and intercommunity dyadic strength and neighborhood overlap, we define a typology of *migration community signatures*.

*Cave communities*

We say that the cluster of communities with negative scores of E-I strength–overlap indices are *cave*[11] communities. For example, 14 cave communities constitute the blue cluster in Fig. 8. The communities centered on Russia and West Africa are examples of cave communities. The Russia-centered community falls under the cave category even when using the spatial null model, suggesting that this is the community that has the highest local cohesion, which is preserved even after we factor out geographic distance, one of the key mechanisms of local connectivity.

Cave communities are characterized by having both low $EI_{es}$ and low $EI_{no}$. The structure of cave communities is therefore characterized by strong local cohesion—i.e., densely clustered, strong migration ties—but weak global cohesion (i.e., lack of bridging weak ties across communities), resulting in tightly knit migration interactions that are largely fragmented from the rest of the WMN.

Given the structure of cave communities, they may be associated with distribution of regional migratory movements, while simultaneously providing very limited opportunity structures for intercommunity migration connectivity. Movements of people that originate from cave communities are largely constrained to remain within communities due to the limited amount of weak bridging edges that channel migration to other communities.

*Biregional communities*

We refer to the communities clustered in the middle of the E-I strength–overlap place as *biregional* communities (see the green cluster, which includes 29 communities, in Fig. 8) because they are neither exclusively local nor exclusively global but rather often connect two distinct regions in the WMN. Examples of biregional communities include the one that connect Arab countries and South Asia (including Asia) and the one that connects France and countries from North Africa. Biregional communities tend to resemble cave communities in $EI_{es}$ but resemble bridging communities in $EI_{no}$ (see Fig. 9). Because biregional communities encompass patterns of both local cohesion and global cohesion, they instantiate tendencies towards glocalization.

---

[11] We draw the notion of 'caves' from Watts (1999) and Martin (2009). In the original 'caveman graph', caves refer to $k$-cliques.

*Bridging communities*

We say that communities that occupy the positive end of the E-I strength–overlap continuum are *bridging* communities. For example, 22 bridging communities constitute the brown cluster in Fig 8. Example bridging communities are the following: the ones centered on France, Germany, the United Kingdom, and the USA. A characteristic feature of bridging communities is the predominance of edges that form bridges between regions in the WMN. As one can see from the circular plots[12] in Fig. 10, the largest bridging community (USA) is simultaneously connected to communities DEU, GBR, IND, and CHN (LN null model), and communities DEU, GBR, IND, and CHN (Spa null model), some of which are relatively disconnected from one another.

      Bridging communities tend to provide better opportunities than cave and biregional communities for both cross-community mobility and cross-continental exchanges, as they often group non-contiguous countries across continents. The classification of the Commonwealths as a bridging community suggests that communities of this type may reflect underlying homophilous relationships, a phenomenon that we explore further below.

---

[12] For a recent use of circular plots to visualize global migration flows between geographic regions, see Abel and Sander (2014).

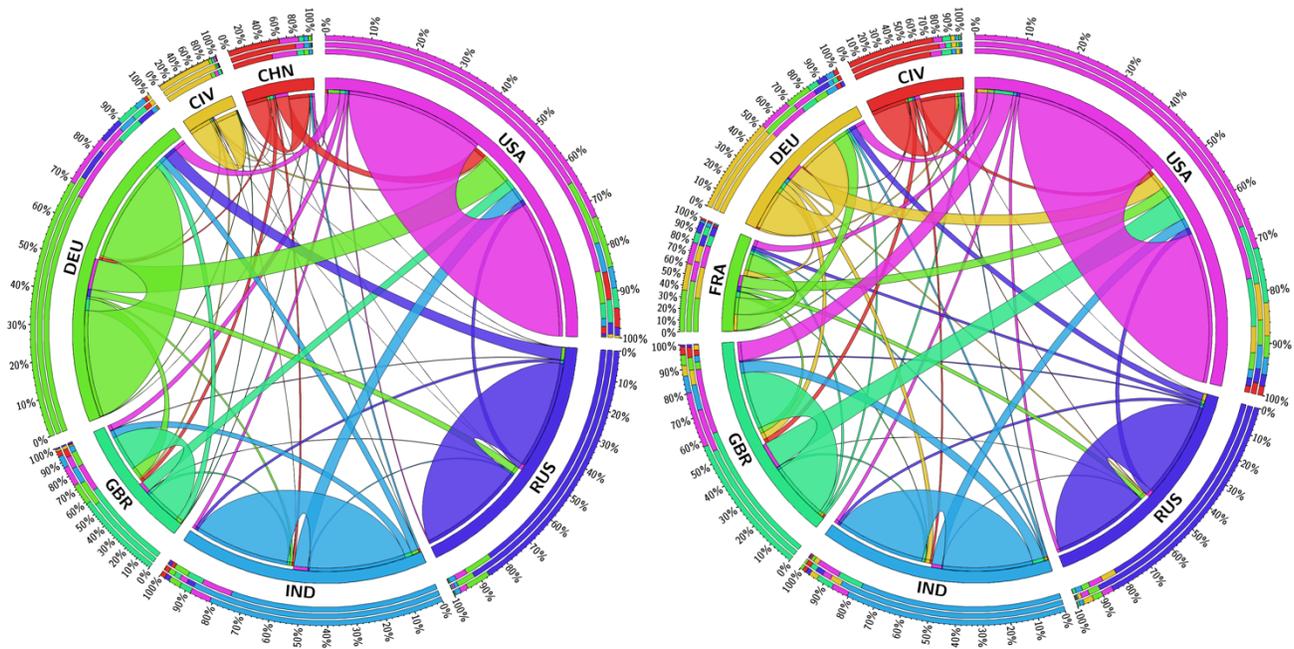

Fig. 10. Circular plots of migration community structures in 2000. The size of the ribbons corresponds to the amount of migration stock that remains in a community or is directed to other communities. The color of the ribbons indicates the source communities. We create the plots using Circos Table Viewer (Krzywinski et al., 2009), which is available at http://mkweb.bcgsc.ca/tableviewer/visualize/.

## 7. Continuity and Change in Migration Communities

Migration communities are involved in complex processes of emerging, splitting, merging, and dissolving. In Fig. 11, we map continuity and change in migration communities using alluvial diagrams (Rosvall and Bergstrom, 2010). Instead of processes of integration, we observe a split in bridging communities since 1960, with a noticeable effect in the last decade. However, there are also instances of merging of newly industrialized areas, e.g., the CHN community in 1960 joins the largest bridging community in 2000 (spatial null model). Such processes of consolidation are likely to result from direct foreign investments and manufacturing export, which induce relationships between distant regions (e.g., North America and South Asia) in a network of socio-economic relationships (Sassen, 2007, Castells, 1996). Cave communities (e.g., RUS, CIV) are relatively isolated from the temporal dynamics in the WMN.

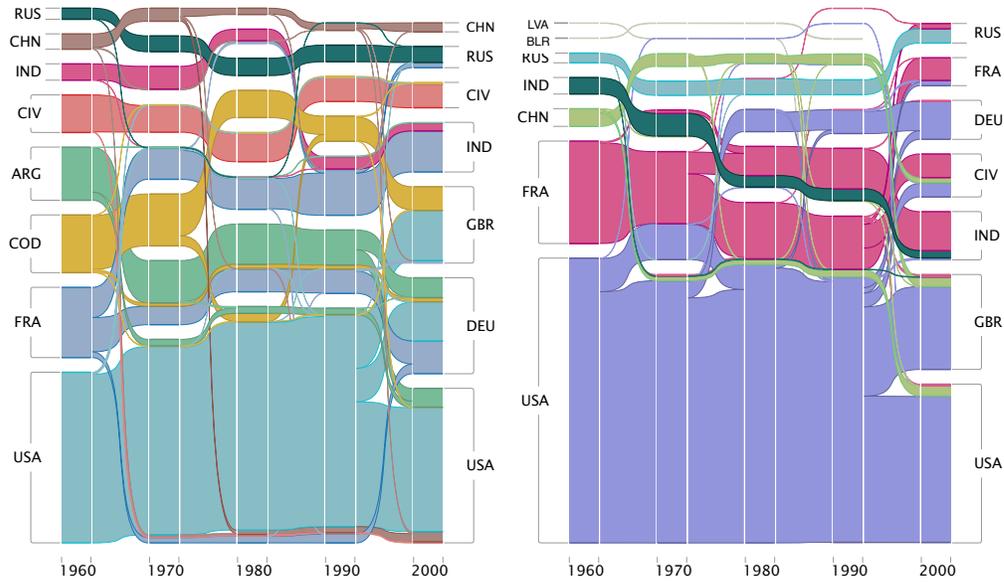

Fig. 11. Temporal changes in migration communities that we detect by maximizing modularity with (left) the LN null model and (right) the spatial null model.

To examine processes of continuation and change in the membership structure of migration communities, we compute community autocorrelation $C(t)$ (Palla et al., 2007). Cave communities have a very stable structure $C(t) \approx 0.93$, so on average fewer than one out of ten countries changes community membership from one time point to another. Simmel's (1950[1908]) observation once social structures tend to have a life of their own once they emerge from social interactions is therefore of particular relevance to cave communities. By contrast, the countries in biregional and bridging communities in particular change much more. The estimated means for those community signatures are $C(t) \approx 0.61$ and $C(t) \approx 0.67$, respectively. An ANOVA multiple-comparisons test estimates that the autocorrelation means of cave communities are significantly different from those of bridging and biregional communities. The means of biregional and bridging communities are not significantly different.

The size $N_c$ of communities may be a confounding variable. As Palla et al. (2007) determined, large communities tend to have a higher rate of change compared to small ones. We find limited ($r \approx -0.18$) evidence (of border significance level, $p$-value $\approx 0.1$) in support of the hypothesis that community size $N_c$ and community stability $C(t)$ are correlated. We find stronger and significant ($p$-value $\approx .005$) negative correlations between $C(t)$ and $EI_{es}$ ($r \approx -.4$) and between $C(t)$ and $EI_{no}$ ($r \approx -.43$). Community stability therefore decreases with an increase of global cohesion.

Our results do not fully support the observation that "the legacy of old communities tends to disappear in time" (Davis et al., 2013). Although $C(t)$ scores differ somewhat across decades (particularly in 2000), neither a classic ANOVA test nor a non-parametric Kruskal–Wallis test found those differences to be significant. In other words, the mean (and median) differences in community autocorrelation in 1970, 1980, 1990, and 2000 do not differ significantly (1960 is a reference year against which we compute $C(t)$ for 1970). This finding has two implications. First, differences in the WMN appear to be more pronounced across spatial network community signatures than across time. Second, the structure of the WMN over the second half of the 20th century exhibits historical continuity.

## 8. Relational, Homophily, and Spatial Antecedents

What antecedents could have brought about the heterogeneity of migration community signatures? Drawing upon the insight that collectives that differ in macroscale properties will also differ in their microscale properties (Hedström and Bearman, 2009), we hypothesize that migration communities with distinct signatures are likely to arise from different underlying mechanisms.

### 8.1. Community Signatures in Multidimensional Space

We employ PCA (Jolliffe, 2002) to arrange our set of relational, homophily, and spatial mechanisms in a multidimensional space (see Fig. 12). Individual migration communities occupy particular locations in this space depending on how they are associated with one or another mechanism. We consider the first three principal components because they account for 78% of the total variation among the battery of community diagnostics.

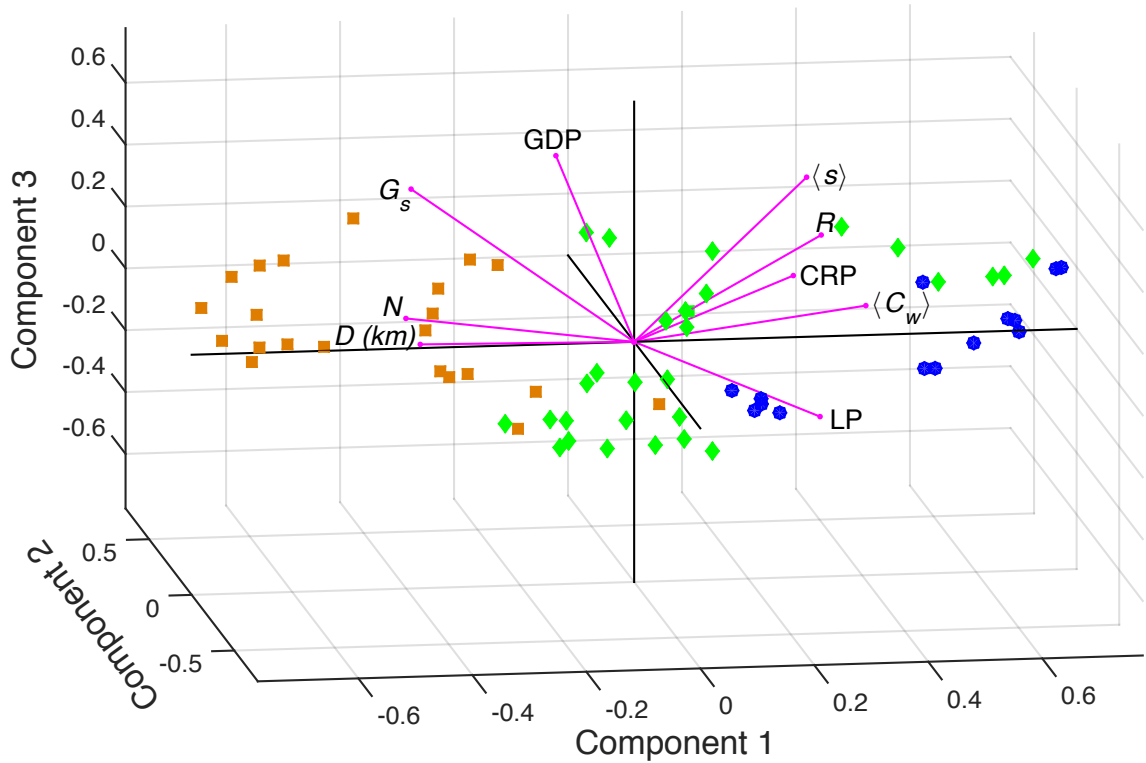

Fig. 12. Three-component PCA. We show migration community signatures with blue dots (cave communities), green diamonds (biregional communities), and brown squares (bridging communities). Each vector corresponds to one of the original community diagnostics, and the length of the vector indicates the strength of the contribution.

We find that communities with a similar migration signature—i.e., a similar pattern of local and global cohesion—appear near each other in the three-dimensional principal component space. Further, different community signatures are arranged around different diagnostics. Bridging communities tend to sit together in the upper left corner, aligning with unequal strength distribution (measured via $G_s$), size, and distance. Cave communities are located predominantly in the lower right in association with clustering coefficient ($C_w$) and language homophily. Biregional communities occupy the middle ground, aligning with dyadic forces (reciprocity and ex-colonial homophily), associated with cave communities, and GDP per capita, associated with bridging communities.

Do community signatures differ significantly in the relational, socio-economic, and spatial mechanisms under study? A post-ANOVA multiple-comparisons test establishes that mean differences between each pair of community signatures are significant for three diagnostics: weighted clustering coefficient, reciprocity, and expected distance (see Fig. 13). For the clustering coefficient and reciprocity, cave communities have the largest scores (bridging communities have the lowest). Biregional communities occupy a position in the middle. Further, we observe substantial differences in the expected distance between an origin and a destination for a migrant selected uniformly at random. This distance is about 1,500 km for cave communities, about 2,500 for biregional communities, and about 4,200 for bridging communities. Additionally, we have a statistically significant difference between the mean scores of bridging and cave communities in virtually all diagnostics (chain migration is the only exception), suggesting a pattern of polarization. Bridging communities differ systematically from biregional communities in all variables except ex-colonial and language homophily.

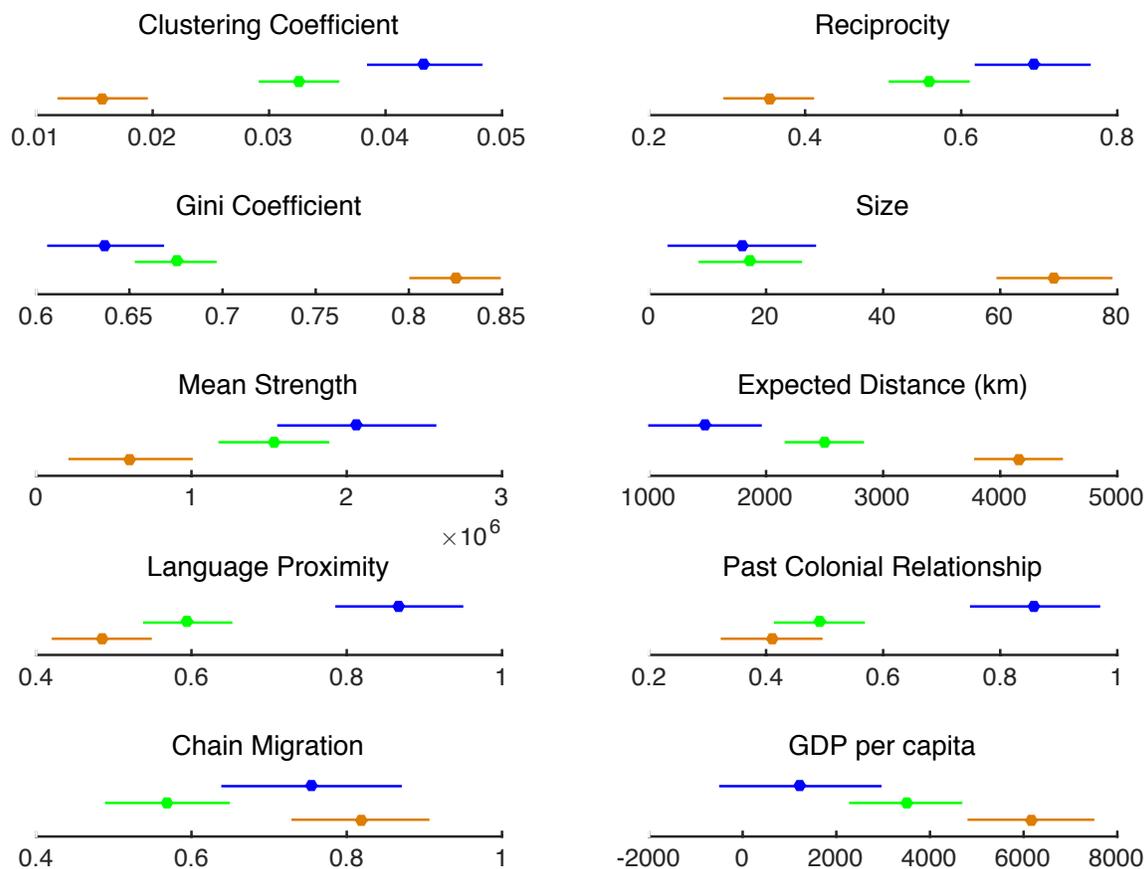

Fig. 13. Multiple comparisons of mean differences in network, socio-economic, and spatial diagnostics between community signatures. In each plot, we show community signatures for (top) cave, (middle) biregional, and (bottom) bridging communities. The lines indicate the interval of standard error, and the symbol in the middle of the interval indicates the mean.

An important global property of bridging communities is their greater inequality in strength distribution: if cave communities have a mean Gini coefficient of $G_S \approx 0.64$, a score value that is similar to that of biregional communities ($G_S \approx 0.67$), bridging communities reach a score of $G_S \approx 0.83$. The $G_S$ cores of cave and bridging communities have polarized. The strength distribution in bridging communities becomes less unequal over the decades (e.g., $G_S$ for the USA community is about 0.85 in 1960 and about 0.77 in 2000). At the same time, $G_S$ of the West African cave community surrounding Ivory Coast increases from $G_S \approx 0.48$ in 1960 to $G_S \approx 0.57$ in 2000. This tendency of change in opposing directions suggests that processes of polarization in the WMN are not an atavistic manifestation, as they also arise in the context of globalization.

## 8.2. Relative Importance of Antecedents Shaping the WMN: MR-QAP

What is the relative importance of migration antecedents? To address this question, we perform a MR-QAP (Krackardt, 1987, Dekker et al., 2007) using weighted migration stock matrices for the WMN, cave (RUS), biregional (IND), and bridging (USA) communities detected via LN modularity as response variables and a set of predictor matrices representing our relational, homophily, and spatial antecedents.

We summarize our results here, and provide details about the model in Appendix B. First, our predictors tend to better explain variations in migration when applied at the scale of communities rather than the whole WMN, suggesting that community-scale MR-QAP regression accounts for heterogeneity in world migration. Second, the effects of relational antecedents on community signatures remain mostly significant after we control for homophily and spatial constraints. Relational effects are stronger for the bridging USA community and for the whole WMN than for the communities centered on India and Russia. Third, the MR-QAP model supports the hypothesis that cave communities are more likely to be affected by multiple spaces (language homophily and spatial proximity) while the bridging USA community follows either/or logic—it could be involved in either relational, homophily, or geographic space. Finally, our predictors provide limited understanding of the evolution of the biregional IND community, merging the Gulf region with countries in South Asia (e.g., India) since 1990. The main reason should be the lack of global data on the role of states and migration policies (Zolberg, 1999). And while migration policies may often align with social and geographic proximity, the Gulf countries since 1970s recruited deliberately migrants from distant—in social and geographic space—areas (Myron, 1982, Massey et al., 1998: 134–159). For this reason, our understanding of the evolution of this migration community is limited. Despite this limitation, the MR-QAP results highlight the heterogeneous structure of the WMN, such that antecedents that may have very little impact on one community signature (or the WMN as a whole) could have a distinct and strong impact on other community signature.

## 9. Conclusion

The foregoing analysis suggests that multiple transnational movements of people across the world, connecting dispersed countries at various geographic scales, have crystallized into heterogeneous—"cave", "biregional", and "bridging"—signatures with distinct patterns of global and local tendencies, temporal dynamics, underlying antecedents, and opportunity structures for future migration.

Compared to related global flows of tourism (Belyi et al., 2016) and trade (De Benedictis and Tajoli, 2011) that have been reported to significantly increase interconnectivity over the latter decades of the 20th century, our findings suggest that world migration is neither globally interconnected nor reproducing the geographic map of the world. The WMN exhibits rather heterogeneous structure, with relatively well-defined and enduring migration community signatures. The increase interconnectedness of the WMN, also reported in previous studies (Fagiolo and Mastrorillo, 2013, Davis et al., 2013) and underpinning the globalization hypothesis, is concentrated in bridging communities. Possibly because of legal restrictions associated with long-distance movements (Massey, 1999), global connectivity does not diffuse into cave communities, which are relatively isolated from the rest of the WMN. Therefore, we observe a polarization tendency within the WMN, with migratory movements in bridging and cave community signatures manifesting opposing patterns and antecedents, which are not just a reminiscent from the past but continue also in the latter decades of the 20$^{th}$ century. Biregional communities instantiate glocalization tendencies as global and local movements coexist in the same spatial environment.

Our findings help to rethink key assumptions of current thinking about globalization. Globalization theories typically construe networks as emerging from the intensification of worldwide interconnectedness of markets, transportation, and communication, which 'cut across the boundaries of the national state' (Beck, 2000: 4) and compress the world to a single place, the globe (Giddens, 1990, Robertson, 1992, Castells, 2010). Our results, however, suggest that the WMN involves heterogeneous sub-structures, only some of which are interconnected in line with the globalization argument. *Globalisation* is *local* in a sense that global processes are specific to some regions of the network but do not operate in others. This is probably why globalization tendencies in world migration have not translated to a relatively integrated global labour market (Hirst and Thompson, 1999: 275). Certainly, globalisation scholars have already pointed to the variability of connectivity across localities (Dicken et al., 2001: 96). Our findings provide not only an empirical support of such suspicion but also a quantitative characterization of variations in connectivity.

Globalization theory conceptualizes networks as ontological forms that manifest a new type of liberating social organization, a "network society" (Castells, 2010), which is structured around "space of flows" replacing the hierarchical and bureaucratic structures of bounded nation-states ("space of places"). However, it has long been established that real-world network structures may not only enable given outcomes but could also constrain action or flows (Wellman and Berkowitz, 1988, Wasserman and Faust, 1994, Borgatti et al., 2009), a property that is particularly evident in fragmented networks (González-Bailón and Wang, 2016). Likewise, migration community signatures can have not only a liberating effect but also important constraining implications hampering migration opportunities.

The world systems theory (Wallerstein, 1974) accounts for polarization and constraining effects arising endogenously from increasing interdependencies in global economy. Our results provide little support for the hypothesis that processes of polarization in the WMN are a function of economic dependencies rather than isolation. The most disadvantaged communities appear isolated, either politically or economically. At the same time, regions in Asia have become more interdependent into the world economy since 1970s when core capitalist countries (e.g., USA) relocated labour-intense manufacturing to periphery countries via foreign direct investments (Sassen, 2007: 36–37). This formed global economic links that eventually channelled long-distant migration (Sassen, 2007). As a result, contrary to the world system theory prediction, economic interdependence has been translated into greater migration opportunities.

Globalization has often been viewed as a 'new historical conjuncture' (Glenn, 2007: 34, McGrew, 1998). By contrast, our investigation of spatial network community signatures in the WMN highlights processes of historical continuity. Although the WMN has indeed become more interconnected as a whole, the migration community signatures have not changed significantly over time. Cave communities are as much as isolated from the network in 2000 as in 1960, an example of polarization. We acknowledge, however, that both our data (aggregate migration stocks) as well as methodology (community detection in spatial and temporal networks) favours continuity at the expense of change.

Our analysis can be extended in several ways. First, upon data availability, one could stratify the edges in the WMN by type of migration (e.g., highly skilled professionals, workers, students, refugees, and family unification) and construct a multilayer network (Kivelä et al., 2014) in which countries are connected via multiple types of migration. A network of highly skilled professionals would generate a different mapping of world migration compared to a network of workers, for example. Such an approach could shed light on the differential impact of migration policies on different types of migration. Second, given the multilateral and multiscale nature of migration exchanges in the WMN, countries can belong to more than one community, pointing to the importance of methods that can discover overlapping communities (Gopalan and Blei, 2013). Third, our approach of tracing the interplay between local and global interactions will broaden if multiple flows of people, information, goods, and capital are considered. Recent studies (Belyi et al., 2016, Hristova et al., 2016) consider some of these flows and could provide a good platform for exploring heterogeneity in a wide range of global processes. Finally, the ubiquity of online information in the public domain provides an opportunity to collect data about human mobility, e.g., geolocated career records in LinkedIn (State et al., 2014), and thus redraw the map of local and global connectivity in world migration using self-reported instead of administrative data. To sum up, our approach of investigating heterogeneous—globalization, polarization, and glocalization—world processes can be extended to multiple flows and data sources, and thus shed light on emerging transnational patterns of migration interactions and possibilities.

Acknowledgements

We thank Basak Bilecen, Adam Dennett, Markus Gamper, Bernie Hogan, Michael Keith, Miranda Lubbers, and Martin Ruhs for helpful comments. We thank Paul Expert for sharing his code for the spatial null model, and we thank Marya Bazzi, Lucas Jeub, Mikko Kivelä, and Marta Sarzynska for useful discussions and help with MATLAB. Both authors were supported by research award (No. 220020177) from the James S. McDonnell Foundation. VD was also supported by the European Community's Seventh Framework Programme (FP7/2007-2013)/ERC grant agreement No. 240940. MAP was also supported by the European Commission FET-Proactive project PLEXMATH (Grant No. 317614).

Appendix A: Consensus Partitions

We employ the technique of consensus partitions (Lancichinetti and Fortunato, 2012, Bassett et al., 2013) to identify robust communities (i.e., communities that do not change substantially from run to run of the heuristic) and minimize the issue of near-degeneracy (Good et al., 2010) of the modularity function. The technique involves the following steps, as performed in Bassett et al. (2013: 13-14), Bazzi et al. (2015), and Sarzynska et al. (2015). First, we construct a new multilayer co-association tensor **T**, which includes the five migration matrices as individual layers. Each element of the tensor $T_{ijl}$ represents the number of times a country $i$ is assigned to the same community as country $j$. We performed the procedure at resolution $\gamma = 1$ for both the LN null model (Leicht and Newman, 2008) and the spatial null model (Expert et al., 2011). Consider an example of 100 partitions, in which country $i$ and $j$ are assigned eighty times to the same community in a given layer. The corresponding element in the tensor would be $T_{ijl} = 80$ and will appear in red in Fig. S1. Second, given the large number of possible pairwise associations in **T**, one needs to account for the probability of two countries being assigned to the same community by mere chance. In order to reduce the noise that could arise from possible false positives, we subtract from **T** the mean number ($\mu \approx 11$) of co-associations in the tensor, referred to as a uniform null model (Bazzi et al., 2015).

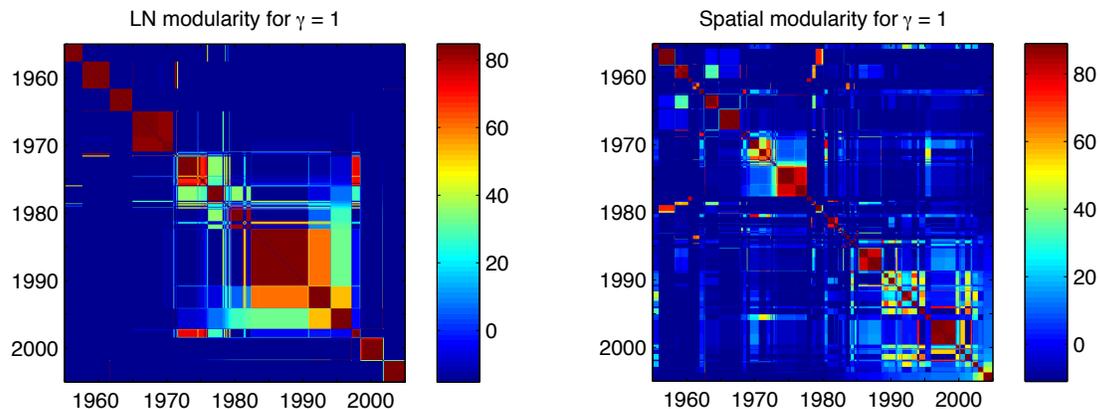

**Fig. S1. Multilayer co-association tensors for the five time periods.** We construct the co-association tensors on the basis of the numbers of times a pair of countries is assigned to the same community in 100 partitions obtained using the LN modularity (left) and spatial modularity (right) at resolutions of $\gamma = 1$. Each tensor has five layers, corresponding to the five migration matrices from 1960 through 2000. The tensors are symmetric. Red colour indicates high association between a pair of nodes, whereas blue colour represents low association. High association in this context indicates that a pair of countries was assigned roughly more than 60 modularity maximisations (out of 100).

The resulting co-association tensors have a characteristic pattern: certain pairs of countries are almost always assigned to the same community (dark red), while the remaining pairs of countries almost never appear in one community (dark blue) (see Fig. S1). High association indicates that a pair of countries appears in the same community because of their migration connectivity rather than as a side effect of the different runs of the algorithm. We subsequently use the tensors as an input to generalised Louvain heuristic instead of the original migration matrices. We generate in this way consensus communities, in which highly associated countries, marked in yellow-to-red, are more likely to form part of the same community. In contrast to the original partitions, which differed from run to run, the consensus partitions are virtually identical across multiple runs at the same resolution $\gamma$ level.

Appendix B: Relative Importance of WMN Antecedents using MR-QAP

MR-QAP
We examine the relative importance of migration antecedents using multivariate-regression quadratic assignment procedure (MR-QAP) (Krackardt, 1987, Dekker et al., 2007). MR-QAP is a multivariate linear regression technique that adopts a non-parametric procedure for testing statistical significance called the Quadratic Assignment Procedure (QAP) (Hubert, 1987). The MR-QAP is tailored to the dependencies in network data, such that the assumption of independence between observations—which is built in the classical parametric statistical significance tests—is not required (Krackhardt, 1988, Borgatti et al., 2013: 126–129). Several permutation techniques for multivariate regression have been proposed (e.g., Krackhardt, 1988). We used the semi-partialling-permutation method that was developed in Dekker et al., (2007) and implemented in UCINET version 6.487 (Borgatti et al., 2002). The method has been reported to be more robust against correlations between the predictor variables (multicollinearity) (Dekker et al., 2007).

Dependent and Independent Matrices
We use weighted directed matrices of world migration stock for the WMN, cave (RUS), biregional (IND), and bridging (USA) communities at a given year as the response variable. Recall that a weighted matrix is defined as follows: an edge represents the number of migrants from a receiving country $i$ residing in a receiving country $j$ at a given decade. An edge does not exist (i.e., it is '0') if there was no migration between the pair of countries. We took the natural logarithm of each element in the weighted matrix of world migration to control for positive skewness. We apply the same transformation to the other weighted matrices: distance and GDP per capita.

    We examine the effect of relational mechanism on the community structure of world migration by including reciprocity $R$ and geodesic edge betweenness centrality $C_{EB}$ as indicators of local and global cohesion, respectively. We construct reciprocity matrix by transposing the original directed matrix, as described in Borgatti et al. (2013: 132). To construct the edge-betweenness centrality matrix, we employ an algorithm proposed in Brandes (2001) and implemented in MATLAB (Rubinov and Sporns, 2010). The betweenness centrality $C_{EB}$ of the edge $E$ is defined as the sum of the fraction of all shortest paths in the network that pass that edge. Edges that are involved in a large number of shortest paths gain higher betweenness centrality scores. Edge-betweenness centrality measures the extent to which an edge contributes to the global connectivity of the WMN.

    We include two variables—language proximity and former colonial relationships—to measure homophily effects. In the matrix of language proximity we construct, '1' signifies that country $i$ and country $j$ share the same official language or that at least 9% of the population in the dyad of countries speak the same language. Otherwise, the matrix element is set to '0'. We construct the matrix of common colonial past in a similar fashion. We place a '1' if two countries have ever had a colonial link or have had a colonial relationship since 1945, and we otherwise place a '0'. We use CEPII Geodesic Distance Database (Mayer and Zignago, 2006).

    To control for spatial effects, we use two variables: geographic proximity and

contiguity (common border). We compute geographic proximity as the great-circle distance (in kilometres) between the capital cities in country $i$ and country $j$, using the package 'fields' in R (Furrer et al., 2013). Recent studies on online social networks have utilised data on the frequencies of airline flights between pairs of places, thereby providing a more realistic approximation of socio-economic costs associated with spatial disparities compared to geographic distance per se (Takhteyev et al., 2012). However, historical data at a global scale is not available. In the contiguity matrix, we define as '1' if country $i$ and $j$ share a border and '0' otherwise. Finally, we define the GDP per capita matrix as the log difference between the GDP per capita of country $i$ and country $j$.

Results

Our predictors tend to better explain variations migration when applied at the scale of communities rather than the whole WMN, as indicated by the larger adjusted $R^2$ coefficients for migration communities, and the bridging USA community in particular (see Table S1). This is possibly because community-scale MR-QAP regression accounts for heterogeneity in world migration.

Relational effects on community signatures are mostly significant after we control for homophily and spatial constraints, although their distribution is unequal. Relational effects are stronger for the bridging USA community and for the whole WMN than for the communities centered on India and Russia. For example, there is a significant negative relationship between geodesic edge betweenness centrality and migration edge strength for the whole WMN and the bridging USA community. This is consistent with the strength-of-weak-ties hypothesis (Granovetter 1973). The effect of edge betweenness is not significant for the cave and biregional communities that we study, suggesting that migration exchanges related to those communities contribute less to global connectivity.

The impact of homophily mechanisms on migration exchanges tends to be significant at both network and community scales. The impact of ex-colonial relationships (and associated socio-cultural similarity) is higher than the impact of language similarities. This finding is consistent with the results reported in Mayda (2010). However, the effect of language is strong and significant for the cave community centered on Russia in 1960. It is even larger in 2000.

The impact of economic disparities is pronounced in the bridging USA community, where migrants tend to prefer destinations with higher GDP per capita. The asymmetric movements from spokes to hubs in bridging communities are therefore explained partly by economic differences. By contrast, economic differences play a less important role in the WMN as a whole (particularly in 1960) and in cave and biregional communities. As spatial interaction models predict (Wilson and Oulton, 1983), movements between distant origin and destinations are associated with an expectation for greater economic differentials than one obtains from small-distance movements.

|  | Predictors | WMN | Cave RUS | Biregional IND | Bridging USA |
|---|---|---|---|---|---|
| **1960** | *Relational* | | | | |
| | Reciprocity | 0.418*** | 0.032 | 0.271* | 0.478*** |
| | | (0.015) | (0.032) | (0.153) | (0.047) |
| | Geodesic betweenness | −0.948*** | 0.137 | 0.491 | −0.806*** |
| | | (0.092) | (0.135) | (0.836) | (0.195) |
| | *Social* | | | | |
| | Ex-colonial Relationship | 1.725*** | 1.359*** | 0.000 | 1.305*** |
| | | | (0.466) | | (0.388) |
| | Language Proximity | 0.062 | 0.880* | −1.287 | 0.014 |
| | | (0.044) | (0.406) | (2.06) | (0.139) |
| | *Economic* | | | | |
| | Log (GDP per capita) | 0.001*** | | −1.258* | 0.284** |
| | | (0.0001) | | (0.905) | (0.103) |
| | *Spatial* | | | | |
| | Log (Distance) | −0.329*** | −0.268** | 0.797 | −0.125 |
| | | (0.050) | (0.376) | | (0.124) |
| | Contiguity | 1.313*** | 0.747 | 2.781* | 0.845 |
| | | (0.159) | (0.528) | (1.840) | (0.634) |
| | (Intercept) | 6.211*** | 11.013*** | 2.795*** | 3.106*** |
| | Observations (dyads) | 7597 | 238 | 28 | 580 |
| | Adj. $R^2$ | 0.406 | 0.406 | 0.551 | 0.447 |
| **2000** | *Relational* | | | | |
| | Reciprocity | 0.411*** | 0.175** | 0.051 | 0.515*** |
| | | (0.011) | (0.071) | (0.050) | (0.030) |
| | Betweenness | −1.213*** | −0.216 | −1.696*** | −0.776*** |
| | | (0.063) | (0.146) | (0.497) | (0.181) |
| | *Social* | | | | |
| | Ex-Colonial Relationship | 1.527*** | 2.025*** | 0.000 | 1.031* |
| | | (0.129) | (0.538) | (0.000) | (0.453) |
| | Language Proximity | 0.154*** | 0.996** | 0.546* | 0.414*** |
| | | (0.035) | (0.385) | (0.280) | (0.074) |
| | *Economic* | | | | |
| | Log (GDP per capita) | 0.147*** | –0.107 | –0.001 | 0.233*** |
| | | (0.013) | (0.117) | (0.122) | (0.042) |
| | *Spatial* | | | | |
| | Log (Distance) | −0.403*** | −0.695** | −0.700*** | 0.167* |
| | | (0.032) | (0.278) | (0.270) | (0.083) |
| | Contiguity | 1.409*** | 0.358 | 1.430** | 1.459*** |
| | | (0.119) | (0.462) | (0.552) | (0.299) |
| | (Intercept) | 5.996*** | 13.019*** | 12.617 | −0.539 |
| | Observations (dyads) | 20039 | 272 | 506 | 3422 |
| | Adj. $R^2$ | 0.406 | 0.471 | 0.236 | 0.519 |

*p < .05, **p < .01, ***p < .001.
Table S1. Results of MR-QAP regression analysis of the WMN and selected migration communities for years 1960 and 2000. Standard errors are shown in parentheses. *Note:* the GDP data for community RUS are missing for 1960.

The MR-QAP model supports the hypothesis that cave communities—e.g., the community centered on Russia—are affected by multiple spaces. Both the effects of language homophily and spatial proximity are larger in 2000 than in 1960. By contrast, the bridging USA community follows either/or logic—it could be involved in either relational, homophily, or geographic space.

Our model is inconclusive with respect to the biregional community centered on India. The community enlarges from 8 countries in 1960 to 23 countries in 2000, when it also includes oil-producing countries from the Persian Gulf (Bahrain, Kuwait, Oman, Qatar, and the United Arab Emirates). However, our model seems to account insufficiently for the mechanisms that have governed community evolution, as one can see from the low adj. $R^2 = 0.236$ in 2000 compared to adj. $R^2 \approx 0.551$ in 1960. One possible reason is that our model does not consider the role of states and migration policies (Zolberg, 1999) because of the lack of comparative and longitudinal data for all 226 countries. A body of literature, which discusses the proactive migration policies of the governments in the oil-producing states that surround the Persian Gulf since 1970s (Myron, 1982, Massey et al., 1998: 134–159), suggests that policies in the region aimed to attract short-term migration from distant—in social and geographic space—countries in South Asia and simultaneously restricted the entrance of migrants from geographically and socially close fellow Arab countries like Egypt and Lebanon. Inasmuch as policies successfully shifted migration patterns away from geographic and social proximity, our model provides a limited understanding of the processes that underlie the transformation of this community.